\documentclass[conference]{IEEEtran}
\IEEEoverridecommandlockouts
\usepackage{cite}
\usepackage{amsmath,amssymb,amsfonts}
\usepackage{algorithmic}
\usepackage{graphicx}
\usepackage{textcomp}
\usepackage{xcolor}
\usepackage{subfiles}
\usepackage{subcaption}
\usepackage{float}
\usepackage{hyperref}

 \usepackage[final]{microtype}
 \usepackage[italic]{mathastext}
 \usepackage{libertine}
 \usepackage[T1]{fontenc}
 \usepackage[varqu,varl]{zi4}
 \usepackage[all]{nowidow}
 \usepackage[keeplastbox]{flushend}
 \usepackage{fancyhdr}
 \usepackage[skins]{tcolorbox} 
 \usepackage{booktabs}
\def\BibTeX{{\rm B\kern-.05em{\sc i\kern-.025em b}\kern-.08em
    T\kern-.1667em\lower.7ex\hbox{E}\kern-.125emX}}
\begin{document}

\title{Belenos: Bottleneck Evaluation to Link Biomechanics to Novel Computing Optimizations
}

\author{\IEEEauthorblockN{Hana Chitsaz}
\IEEEauthorblockA{\textit{Computer Science} \\
\textit{University of Maryland}\\
College Park, MD, USA \\
hchitsaz@umd.edu}
\and
\IEEEauthorblockN{Johnson Umeike}
\IEEEauthorblockA{\textit{Computer Science} \\
\textit{University of Maryland}\\
College Park, MD, USA \\
jumeike@umd.edu}
\and
\IEEEauthorblockN{Amirmahdi Namjoo}
\IEEEauthorblockA{\textit{Computer Science} \\
\textit{University of Maryland}\\
College Park, MD, USA \\
namjoo@umd.edu}
\and
\IEEEauthorblockN{Babak N. Safa}
\IEEEauthorblockA{\textit{Medical Engineering} \\
\textit{University of South Florida}\\
Tampa, FL, USA \\
babakn@usf.edu}
\and
\IEEEauthorblockN{Bahar Asgari}
\IEEEauthorblockA{\textit{Computer Science} \\
\textit{University of Maryland}\\
College Park, MD, USA \\
bahar@umd.edu}
% \and
}

\maketitle

\begin{abstract}

% This document is a model and instructions for \LaTeX.
% This and the IEEEtran.cls file define the components of your paper [title, text, heads, etc.]. *CRITICAL: Do Not Use Symbols, Special Characters, Footnotes, 
% or Math in Paper Title or Abstract.
Finite element simulations are essential in biomechanics, enabling detailed modeling of tissues and organs. However, architectural inefficiencies in current hardware and software stacks limit performance and scalability, especially for iterative tasks like material parameter identification. As a result, workflows often sacrifice fidelity for tractability. Reconfigurable hardware, such as 
FPGAs, offers a promising path to domain-specific acceleration without the cost of 
ASICs, but its potential in biomechanics remains underexplored. This paper presents Belenos, a comprehensive workload characterization of finite element biomechanics using FEBio, a widely adopted simulator, gem5 sensitivity studies, and VTune analysis.
VTune results reveal that smaller workloads experience moderate front-end stalls, typically around 13.1\%, whereas larger workloads are dominated by significant back-end bottlenecks, with backend-bound cycles ranging from 59.9\% to over 82.2\%.
Complementary gem5 sensitivity studies identify optimal hardware configurations for Domain-Specific Accelerators (DSA), showing that suboptimal pipeline, memory, or branch predictor settings can degrade performance by up to 37.1\%. These findings underscore the need for architecture-aware co-design to efficiently support biomechanical simulation workloads.

\end{abstract}


% \note{Babak has edited one and Bahar went through and cleaned up the citations. Almost ready. All, please take a look.}

% \Babak{Could you please help editing this paragraph and suggest sufficient citations to make sure the motivation is solid?}
% \Hana{Add all the citations suggested by Babak below.}

Finite element analysis (FEA) is a cornerstone of computational biomechanics, enabling simulations of the mechanical behavior of complex biological systems under varied conditions. These simulations have wide-ranging applications in healthcare, from patient-specific treatment planning~\cite{merema2021patient, garcia2018step, eggermont2018can} to investigations of tissue and cellular mechanics across scales~\cite{morgan2016finite, feola2016finite, cao2011three, safa2019helical}. Despite their importance, such simulations remain computationally demanding and difficult to scale. System-level inefficiencies in software and hardware force engineers to simplify models through geometric reduction, mesh coarsening, or parameter approximations to achieve feasible runtimes. This reduces model fidelity and constrains iterative workflows such as parameter identification that rely on repeated simulations~\cite{safa2021, safa2020evaluation, ross2024bayesian, lee2021vivo, li2024aging}, ultimately limiting the translational impact of biomechanics, where turnaround time and precision are crucial.

Recent advances in hardware acceleration have greatly improved scientific computing, especially for large systems of equations. Early work targeted GPUs~\cite{micikevicius20093d, michea2010accelerating, giles2014gpu, markall2010towards, balevic2008accelerating, dziekonski2012finite, okimura2013parallelization, fu2014architecting, huthwaite2014accelerated, martinez2015efficient, kopysov2016scalability}, followed by tensor cores~\cite{chen2024convstencil, han2025flashfftstencil, zhang2024lorastencil}, analog and hybrid architectures~\cite{guo2016energy, huang2017hybrid, feinberg2018enabling}, PIM~\cite{chen20201, mu2024scalable, giannoula2022sparsep}, HBM-adjacent memory~\cite{feldmann2024azul}, wafer-scale platforms~\cite{rocki2020fast, jacquelin2022scalable}, and FPGAs~\cite{bakhtiar2024acamar, asgari2020alrescha, li2023fdmax, xu2018improved, gerami2024gust, ramchandani2023spica}. However, their applicability to FEA in biomechanics remains largely unexplored.

A key step toward accelerating biomechanics simulations is understanding how these workloads interact with modern architectures. Existing profiling studies on scientific workloads~\cite{peverelli2022characterizing, gomez2022benchmarking, asgari2021copernicus} do not address the unique computational demands of biomechanics, leaving a gap in architectural characterization. To fill this gap, we present Belenos\footnote{\textbf{Bélénos} is a single star in the equatorial constellation of Pisces. It is also an ancient Celtic deity associated with healing and health.}, which makes the following contributions:

$\bullet~$\textbf{Full-System Performance Profiling:} We provide the first in-depth architectural characterization of FEBio~\cite{maas2012febio,febioGithub,febioStudioUserManual,FEBio_Theory_Manual}, the leading open-source FEA package in biomechanics, using Intel VTune~\cite{intel2023vtune}. By profiling diverse models, we reveal the interplay between compute- and memory-bound phases and identify key bottlenecks.

$\bullet~$\textbf{Ocular Biomechanics Case Study:} We include a glaucoma-focused case study~\cite{safa2023effects}, demonstrating how sparse, data-dependent patterns stress memory hierarchies and exacerbate irregular access.

$\bullet~$\textbf{Architectural Sensitivity via gem5:} Using gem5~\cite{gem5}, we explore how workload performance responds to architectural parameters such as core frequency, cache size, and pipeline width, guided by insights from our real-world system profiling.

By providing the first detailed architectural study of biomechanics simulations, Belenos lays the groundwork for co-designing compute platforms tailored to these workloads. Our analysis shows that targeted architectural enhancements could significantly mitigate current bottlenecks, enabling more realistic and clinically meaningful models. Finally, we examine emerging strategies such as near-memory processing~\cite{ahn2016pim,fujiki2019transformation,gokhale2015rearrangement,ke2020recnmp,kwon2019tensordimm,khoram2017challenges,lloyd2017lookup,rodrigues2019scatter,tanabe2011gather}, lightweight reconfigurable logic~\cite{bakhtiar2024acamar,lin2013design,rodriguez2019blas, yadav2025dynaflow}, and task-specific accelerators~\cite{asgari2022fafnir,asgari2020alrescha,feldmannazul,hegde2019extensor,nurvitadhi2015sparse,pal2018outerspace, vadlamudi2024electra, bakhtiar2024pipirima, hosseini2025segin}, discussing their potential to address these challenges and their practicality for iterative biomechanics workflows.

\section{Motivation}
%add citations
% \note{ready for all to review. Bahar edited once.\\}
% \Bahar{Hana, please add figures}
% \Babak{Could you please help with editing this paragraph and suggest sufficient citations to make sure the motivation is solid?}

\subsubsection*{Why this work} Biomechanical simulations, commonly using FEA, are highly versatile tools for understanding and predicting the physical behavior of biological tissues. However, real-world deployment of these models is often hindered by prohibitive runtimes and system inefficiencies. As a result, biomechanical engineers routinely simplify simulations through geometry reduction, coarse meshing, and simplistic constitutive models, preprocessing steps that compromise model fidelity and physiological relevance. These simplifications are necessary to meet practical computational constraints but limit the complexity and clinical applicability of the models.

%removing for space 
% These limitations directly impact the translational capacity of biomechanical simulations in healthcare. Clinical workflows often demand rapid and accurate simulation results to support treatment planning and decision-making. Yet, current software and hardware designs fall short in meeting this need. Additionally, many biomechanical investigations rely on iterative methods such as parameter identification or optimization, which require repeated simulation runs. The inefficient performance of existing systems makes such iterative pipelines impractical at scale, constraining both research progress and translational adoption.

\subsubsection*{What recent studies have done}

% Despite these capabilities, FEBio’s performance at the system level remains under-characterized. While prior work has validated FEBio’s biomechanical accuracy and compared it with other FEA tools, no study to date has explored how FEBio interacts with modern hardware architectures—particularly with respect to memory access, computational bottlenecks, and workload-specific demands. This presents a gap in the current literature and a unique opportunity for architectural analysis.
To run scientific computing workloads faster and more efficiently, accelerator studies began over a decade ago by targeting GPU-accelerated computation~\cite{micikevicius20093d, michea2010accelerating, giles2014gpu, markall2010towards, balevic2008accelerating, dziekonski2012finite, okimura2013parallelization, fu2014architecting, huthwaite2014accelerated, martinez2015efficient, kopysov2016scalability, micikevicius20093d} with a focus on parallelization and more recently leveraging tensor cores~\cite{chen2024convstencil, han2025flashfftstencil, zhang2024lorastencil}. Later, following the advent of domain-specific architectures (DSAs) for artificial intelligence and neural networks, computer architects realized that these could also be advantageous for scientific computing. As a result, since 2016, various DSAs have been proposed with a focus on analog computing~\cite{guo2016energy}, hybrid analog-digital solutions~\cite{huang2017hybrid}, memristive-based scientific computing~\cite{feinberg2018enabling}, processing-in-memory~\cite{chen20201, chen2019sram, mu2024scalable, feng2024efficient, xie2024fast, giannoula2022sparsep}, leveraging distributed on-chip memory~\cite{feldmann2024azul} near High-Bandwidth Memory (HBM) solutions~\cite{singh2020nero}, and extending wafer-scale machine-learning accelerators stencil computation~\cite{rocki2020fast, jacquelin2022scalable, ltaief2023scaling}. 
%% FROM REBUTTAL
GPU acceleration is another common solution. However, based on existing literature and ongoing work with sparse and scientific computing workload acceleration on Alveo FPGA boards, % ADD CITATION %in our lab 
GPU acceleration tends to underperform on workloads that are not highly parallelizable. Specifically, the sparse, memory-intensive, and branch-heavy nature of the studied workloads results in irregular memory access patterns and complex control flow, which cause branch divergence and non-coalesced memory accesses—issues that significantly degrade GPU performance. 
In contrast, FPGA-based accelerators have shown promise in efficiently handling sparse matrix operations and other scientific computing bottlenecks by enabling more specialized architectures tailored to these irregular workloads%. 
%%%%%%
% FPGA implementations of key kernels used in scientific computing have also been proposed by several studies
~\cite{bakhtiar2024acamar, asgari2020alrescha, li2023fdmax, mu2024scalable, holanda2011fpga, karkooti2005fpga, arias2011suitable, irturk2008fpga, xu2018improved, sun2007implementation}.

While all these methods achieve significant speedups over general-purpose processors, they are often limited to specific algorithms, and their impact on a wider range of problems—even within scientific computing domains—is unclear. Understanding the most efficient hardware solutions for a particular application domain, such as biomechanics, requires specific performance profiling and characterization studies to reveal performance bottlenecks. While several studies have characterized solvers or sparse matrix libraries or accelerators~\cite{peverelli2022characterizing, gomez2022benchmarking, asgari2021copernicus, ozer2025superba}, their findings are either not directly applicable to domains such as biomechanics, or they are performed based on open-source sparse matrix collections which, while reliable, may not represent real-world biomechanics workloads.

%as much as I love this figure I think we need the space 
% \begin{figure}[b]
%   \centering
%   \includegraphics[width=0.8\columnwidth, scale=1]{figs/Belenos.pdf}
%   \vspace{-5pt}
%   \caption{Research goal: bridging modern hardware accelerators with biomechanics applications \Hana{Feel free to remove if space is needed.} %\cite{elmer_pump_image}\cite{tibia_ct_biomechanics}\cite{simonyan2020systemdesign}.
%   }
%   \label{fig:vendiagram}
% \end{figure}

\begin{figure*}[t]
% \vspace{5pt}
\centering
  \includegraphics[width=0.8\textwidth, scale=1]{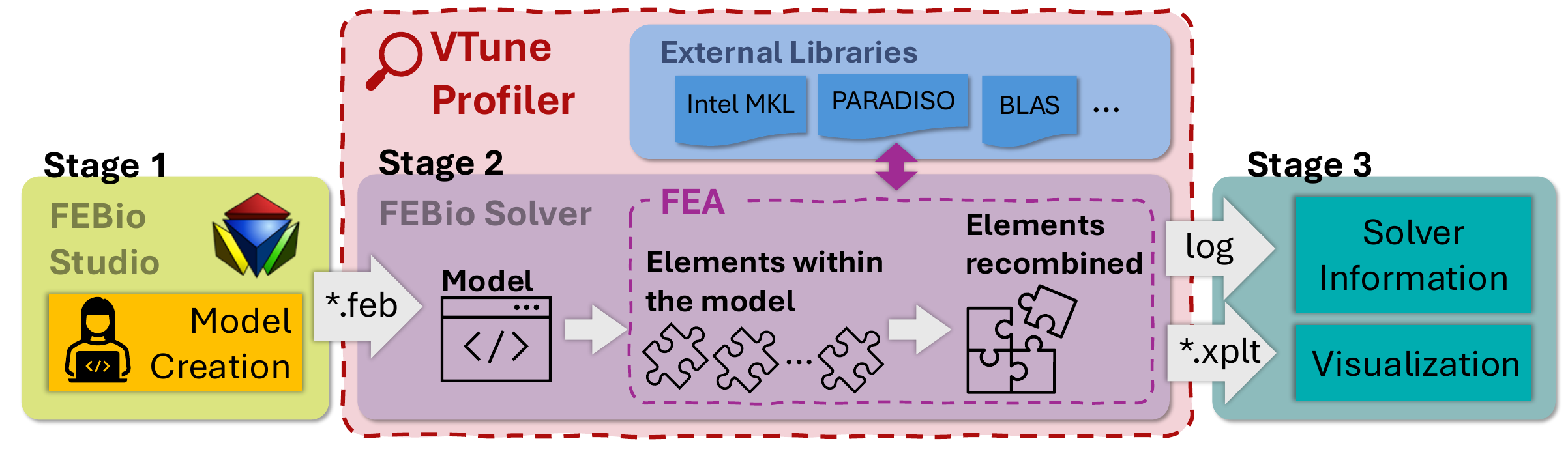}
  \vspace{-7pt}
  \caption{Belenos workflow including the FEBio lifecycle shown in three stages, and the VTune to profile the performance characteristics of Stage 2, which consists of solving the biomechanical models.}
  \label{fig:lifecycle}
  \vspace{-10pt}
\end{figure*}

% \begin{figure}[t]
%   \includegraphics[width=3in]{figs/sources/febioLifecycle.png}
%   \caption{The FEBio Lifecycle.}
%   \label{fig:lifecycle}
% \end{figure}

%\vspace*{-.25cm}
\subsubsection*{Key research questions}This gap presents an opportunity. To improve performance in biomechanics, we must first understand how simulations behave under modern architectures. Therefore, Belenos seeks to answer the following research questions:

    $\bullet~$\textbf{What are the core performance bottlenecks?} Understanding compute, memory, and dataflow limitations reveals opportunities for architectural tuning.

    $\bullet~$\textbf{How does model performance scale with system-level changes?} We explore sensitivity to pipeline width, cache sizes, and other hardware parameters.
 
     $\bullet~$\textbf{Which types of biomechanical models stress the system the most?} Identifying demanding workloads helps target optimizations.

These questions are determined by our initial, inspirational research question: \textit{are there architectural accelerators that match the observed bottlenecks in biomechanics simulations?} 
% We briefly assess the relevance of near-memory processing, reconfigurable hardware, and other architectural paradigms to biomechanics workloads.
%
By characterizing the architectural behavior of biomechanics simulations, we provide a foundation for co-designing hardware and software that better serves the needs of simulation-based science.
%As Figure~\ref{fig:vendiagram} shows, 
Belenos connects the modern architectural strategies to biomechanical simulations to enable more realistic, complex, and efficient biomechanical models, ultimately advancing both scientific discovery and clinical impact.

\section{Background}
%my attempt at shortening the work

To ground our study, Belenos targets Finite Elements for Biomechanics (FEBio) \cite{maas2012febio}, a leading open-source simulation package developed specifically for modeling the mechanical behavior of biological tissues. Designed through collaboration between the University of Utah and Columbia University, FEBio offers various multiphysics capabilities tailored to biomechanics, including large-deformation materials, material anisotropy, multiphasic media, advanced contact modeling, damage, plastic deformation, growth, and fluid-solid interaction \cite{FEBio_Theory_Manual}.
Unlike general-purpose FEA tools, FEBio features a modular architecture, a rich library of nonlinear constitutive models, and extensibility for custom materials, loads, and constraints.
Models are built in FEBio Studio, a GUI environment, and executed by the FEBio solver via file-based command-line interaction. Internally, the solver uses Intel MKL and PARDISO for sparse linear algebra, and OpenMP for multithreading \cite{intel_mkl,intel_pardiso,openmp}. These features make FEBio widely used in academic and clinical biomechanics research \cite{febioGithub,febioStudioUserManual}.
Despite extensive physiological validation \cite{maas2009comparison}, FEBio’s performance characteristics remain understudied from a computer architecture perspective. Belenos addresses this gap by profiling FEBio as a black-box workload under realistic simulation conditions.
FEBio’s simulation workflow consists of three stages (Figure~\ref{fig:lifecycle}):

\textbf{Stage 1-- Model Creation (Preprocessing):} Users construct models in FEBio Studio, a graphical user interface, by specifying geometry, materials, and boundary conditions. The result is a .feb XML input file.

\textbf{Stage 2-- FEA Solver (Core Computation):} FEBio reads the input file, generates the mesh, and assembles a global system of nonlinear equations based on finite element discretization. The solver handles large, sparse stiffness matrices using direct (e.g., PARDISO, Skyline) or iterative (e.g., FGMRES, RCICG) methods, often with optional preconditioning. This stage dominates both runtime and architectural resource use.

\textbf{Stage 3-- Data Export and Visualization (Postprocessing):}  Simulation results are exported for visualization in tools like FEBio Studio or ParaView. Logs include convergence and error data, but offer limited visibility into low-level performance behavior.

Belenos focuses on Stage 2, where architectural inefficiencies, especially around memory access and solver routines, are most acute. By treating FEBio as a black box, our analysis captures end-to-end behavior using system-level profiling tools across diverse biomechanics workloads.

\section{Experimental Methodology}

%potentially reorder
% \note{Bahar read and edited one.}
% \Hana{Could you add a short introductory paragraph for this section to prevent two Headers appear right after each other? \textbf{<- addressed}}
% \Jnote{Section 4 (Methodology), particularly 4.3 and maybe 4.4 appears to me to be detailed and lengthy. Would it be worth considering whether all the VTune analysis steps are necessary for conveying the core goals of the paper? Many readers, especially in this community, might already be familiar with Intel’s Top-Down methodology. Streamlining this part could help with overall readability.  It may be helpful to clarify the rationale for selecting the specific models (e.g., biphasic, material, fluid, those chosen by Professor Babak).}

% \subsection{Machine Specifics}

In this section, we detail our methodology for the experimental results presented in this paper. Experiments are conducted on a workstation configured to reflect a high-performance, commodity-level system accessible to most researchers in biomechanics. The hardware and software environment is as follows:

% \Hana{FEBio version and Linear solver are not a part of "machine specifics" and should not be listed in this sub-section. Please move them. \textbf{<-addressed}}

    $\bullet~$\textbf{Processor}: Intel Core i9-14900K (24 threads, 32 logical processors)

    $\bullet~$\textbf{Memory}: 64 GB DDR5 at 6000 MHz

    $\bullet~$\textbf{OS}: Ubuntu 20.04 LTS (Linux kernel 5.15)

This configuration ensures full utilization of FEBio’s built-in parallelization mechanisms. The solver attempts to parallelize operations such as stiffness matrix assembly and residual vector computation, particularly in models with nonlinear and multiphasic materials. 
%%from rebuttal
All experiments in this study were run multiple times, and we observed minimal run-to-run variation. As a result, the reported values represent stable performance characteristics of the workloads.
%%%%
 While simulations were evaluated with multiple thread configurations, performance gains plateaued beyond eight threads for the test suite workloads. 

%I believe Johnson's comment is addressed now 
%\Jnote{Parallelism is mentioned in a few places, but I didn’t see any corresponding results or plots showing the impact of multiple threads or processes, either on the real system or in gem5. I had previously skipped the core count experiment due to time constraints, and at the time, Hana had observed that increasing thread count did not yield performance gains. Has anything changed since then? Was this aspect explored in the real-system study?}

\begin{table}[t]
\centering
\caption{Dataset Models Breakdown}
\vspace{-5pt}
% \Babak{could you please take a look and let us know if you can replace the KB sizes with mesh size (if it's easy to do)?}
\begin{tabular}{llll}
\toprule
\textbf{Category} & \textbf{Label}  &\textbf{Lower Bound} & \textbf{Upper Bound} \\
\textbf{ } & \textbf{ }  & \textbf{Size (in kB)} & \textbf{Size (in kB)}\\
\midrule

Arterial Tissue & AR & $8.0$ &  $6.37\times10^2$\\
Biphasic & BP & 6.7 & $4.745\times10^2$\\ 
Contact & CO & 5.4 & $3.14\times10^2$\\
% Contact Mechanics Refined & co\_refined & 8704 \\
        Fluid & FL & $1.1  \times10^3$ & $7.4 \times10^3$\\ 
        Muscle & MU & 4.3 & 4.5 \\ 
        Multiphasic & MP & $1.4 \times 10^1$ & $1.374 \times 10^2$ \\ 
        Tetrahedral & TE & 3.7 & $4.31 \times 10^2$ \\ 
        Rigid & RI & $4.7  \times 10^3$ & $4.7 \times 10^3$ \\ 
        Prestrain & PS & $6.4\times 10^3$ & $6.4\times 10^3$ \\ 
        PlastiDamage & PD & 4.9 & 4.9 \\ 
        Multigeneration & MG & $1.784\times 10^2$ & $2.719 \times 10^2$ \\ 
        FSI & FS & $2.15\times 10^1$ & $7.616\times 10^2$ \\ 
        Misc. & MI & $1.1\times 10^3$ & $4.1\times 10^3$ \\ 
        Material & MA & 4 & $6.802\times 10^2$ \\ 
        Damage & DM & 4.7 & $4.602\times 10^2$ \\ 
        Tumor & TU & $6.0\times 10^1$ & $8.3\times 10^1$ \\
        Rigid joint & RJ & 5 & $7.6\times 10^1$ \\
        VolumeConstrain & VC & $2.711\times 10^2$ & $7.345\times 10^2$ \\ 
        BiphasicFSI & BI & $1.5\times 10^3$ & $7.5\times 10^3$ \\ 
        Case Study & Eye  & $9.86\times 10^4$ & $9.86\times 10^4$ \\
\bottomrule
\end{tabular}
\vspace{-10pt}
\label{tab:dataset_models}
\end{table}

\vspace*{-.25cm}
\subsection{Workload Composition}
% \Babak{feel free to modify the below to remove, change, or add more details}
% \Johnson{This needs to be updated with the gem5 workload as well}

We study two categories of FEBio workloads: (1) a diverse set of test suite models spanning 19 categories, consisting of 160 individual test cases, and (2) a high-resolution case study (\texttt{eye}) model shown in Table~\ref{tab:dataset_models}. These workloads vary in size, mesh density, material types, and boundary conditions, capturing a broad spectrum of computational behaviors 
% from rebuttal 
and utilizing multiple commonly used tissue systems.
%%%%%
This diversity enables a comprehensive analysis of architectural bottlenecks across realistic biomechanics simulations.

\paragraph{\textbf{FEBio Test Suite}}  We experiment with the FEBio Test Suite test cases to capture a range of biomechanical behaviors. For VTune analysis on a real system, we focus on three groups comprising 11 test cases that share identical mesh sizes but differ meaningfully in configuration. Group 1 (\texttt{bp07–bp09}) varies anisotropy in hydraulic permeability to evaluate fluid–structure interactions; Group 2 (\texttt{ma26–ma31}) explores different parameterizations of reactive viscoelastic materials; and Group 3 (\texttt{fl33, fl34}) contrasts steady-state and transient fluid simulations. These models are representative in terms of constitutive model complexity, solver behavior, and material response characteristics.
%from rebuttal
They combine multiple scientific domains and add nonlinear, multiphasic, anatomy-derived irregularities that continually change the sparsity pattern, driving backend stalls and degrading cache locality beyond what is seen in single‑domain or other FEA workloads.
%%%%%

For gem5-based microarchitectural sensitivity studies, we select six representative workloads: \texttt{co}, \texttt{ar}, \texttt{dm}, \texttt{rj}, \texttt{tu}, and \texttt{ma}. While the original 11 VTune workloads span diverse behaviors, their memory and compute demands, including cases like the \texttt{eye} model (Ocular Biomechanics Case Study) with a 32\,GB working set, make them infeasible for detailed simulation in gem5. The chosen six enable tractable yet representative analysis, capturing a wide range of computational patterns and architectural stress points.

\paragraph{\textbf{Ocular Biomechanics Case Study}} This simulation models an average human eye using a detailed finite element mesh derived from anatomical measurements~\cite{safa2023effects}. 
% \Hana{[cite PMID: 36745441]}
In slightly more detail, the model examines the biomechanical impact of negative pressure goggles, a novel drug-free treatment for glaucoma, on the corneal and optic nerve head tissues. This model serves as an example of a large, sparse, data-dependent real-world simulation. The model features 
%over 150,000 degrees of freedom,
time and pressure dependent boundary conditions and complex material representations, including nonlinear and incompressible constitutive models. Unlike the compact models in the test suite, the \texttt{eye} model places sustained pressure on memory bandwidth with its size and complexity. % and demonstrates more pronounced backend execution stalls.

% \Hana{Not sure why the following paragraph is here. Either it's relevant to "metrics" or we should completely remove it:}
% For both workload types, we record not only architectural metrics but also relevant biomechanical parameters such as node count, element count, material configuration, and solver tolerances. These parameters allow us to contextualize architectural behavior in terms of biomechanical model design and complexity.

% \subsection{FEBio Setup}
% \note{Do we need this section anymore? Between the background and the machine specifics section, I think we covered most of the details. Please let me know if I'm missing something}
\vspace{-.6em}

\subsection{FEBio \& Profiling Setup}

To systematically characterize the performance of biomechanics simulations, we follow a structured experimental lifecycle consisting of simulation execution, performance profiling, and detailed analysis. 
% Removed because this is a redundant info. You already metioned that.:
% This process is applied across over 150 models from FEBio’s built-in test suite as well as a complex, domain-specific ocular biomechanics case study.
%
Each workload is executed using the FEBio command-line solver while Intel VTune Profiler 2024.1, as shown in Figure~\ref{fig:lifecycle}, records detailed performance data using the Microarchitecture Exploration analysis mode. VTune captures hardware event counters throughout the simulation phases, including stiffness matrix assembly, nonlinear convergence routines, and contact enforcement. While the output simulation files are not used in performance analysis, metadata such as model size, material configuration, and solver parameters are extracted from log files to enrich contextual interpretation.

\vspace{-.6em}
\subsection{Metrics \& Parameters}
% \Amirmahdi{Please check if it includes any parameters you maybe measuring.}
% \Johnson{Please check if it includes any parameters you maybe measuring. (Response) Yes, these all reflect the metrics I am looking into.}

To characterize the architectural behavior of biomechanics simulations, Belenos captures instruction-level efficiency, memory hierarchy usage, and execution bottlenecks across core solver phases. We analyze each simulation through a blend of hardware event counters, source-level correlation, and bottom-up performance traces. VTune’s interface enables identification of hot functions and regions bottlenecked by latency, resource contention, or inefficient memory accesses. Our performance metrics are organized into three main categories:

\subsubsection*{i. Execution Efficiency Metrics}

We capture raw computational cost via CPU time and clockticks, while cycles per instruction (CPI) indicate processor utilization efficiency. Higher CPI values flag stalls or limited instruction throughput. By tracking the percentages of retiring versus speculative execution, we distinguish between productive work and overheads introduced by control-flow mispredictions or pipeline flushes.

% CPU time and clockticks capture the raw computational cost of each workload.

% Cycles Per Instruction (CPI) reflects processor utilization efficiency. High CPI values are indicative of frequent stalls or poor instruction throughput.

% Retiring and speculative execution percentages distinguish useful work from control-flow overhead or pipeline mispredictions.

\subsubsection*{ii. Memory and Cache Behavior}

Cache miss rates at the L1, L2, and LLC levels reveal data locality and reuse characteristics; accompanying memory-bound stall counts and cycle-activity breakdowns quantify penalties from latency and bandwidth bottlenecks. Further attributing stall cycles to specific tiers (L2, LLC, DRAM) pinpoints the depth and primary causes of memory-induced slowdowns.
%addition from rebuttal 
The %representative
\texttt{eye} model, additionally, showed sustained DRAM bandwidth often exceeding 50 GB/s with peaks near 58 GB/s (approaching platform limits –  60 GB/s), and bursty patterns consistent with sparse-stiffness-matrix-assembly and solver phases. 
%%%%%%%%%%%%%%

% Cache miss rates at the L1, L2, and LLC levels provide insight into data locality and reuse.

% Memory-bound stalls and cycle breakdown by activity identify performance penalties due to memory latency and bandwidth constraints.

% Stall cycle attribution by memory tier (L2, LLC, DRAM) highlights the depth and cause of memory-related slowdowns.

\subsubsection*{iii. Instruction Flow and Pipeline Stalls}

We measure serializing instruction stalls to gauge how often data dependencies force serialization, and we account clockticks spent on slow or unresolved operations to expose pipeline bottlenecks and long-latency instructions. Cycle-activity miss statistics further isolate underutilization and scheduling inefficiencies within the instruction pipeline.

% Serializing instruction stalls measure the frequency of operations that prevent parallel execution due to dependencies.

% Clockticks spent on slow or unresolved operations reflect pipeline bottlenecks and long-latency instructions.

% Cycle activity misses help isolate underutilization or inefficient instruction scheduling.

\subsubsection*{Workload Parameters}

Beyond microarchitectural metrics, we annotate each workload with key model-complexity parameters. Element and node counts determine matrix assembly size and memory footprint, while material formulations (e.g., linear elastic, biphasic, multiphasic) drive computational cost and convergence behavior. The richness of boundary conditions, such as contact mechanics or time-varying loads, adds further variability, and solver hyperparameters (tolerance thresholds, maximum iterations) directly influence iteration counts and overall intensity. By correlating these characteristics with architectural metrics, we uncover how biomechanical complexity stresses different hardware subsystems; for example, fine-grained meshes amplify cache pressure, whereas complex material models increase compute-bound execution.
All workloads are profiled under consistent thread affinity and fixed CPU frequency to ensure comparability. A manual review of VTune traces then highlights shared patterns, recurring bottlenecks, and function-level inefficiencies across the simulation set.

% In addition to microarchitectural metrics, we evaluate model complexity as an essential factor influencing architectural behavior. Each workload is annotated with the following parameters.

% Element and node counts, which impact matrix assembly size and memory footprint.

% Material types, such as linear elastic, biphasic, or multiphasic models, which increase computational cost and convergence complexity.

% Boundary condition richness, including contact mechanics or time-varying loads.

% Solver hyperparameters, such as tolerance thresholds and maximum iteration limits, which directly affect iteration counts and overall workload intensity.

% By relating these model characteristics to architectural metrics, we examine how biomechanical complexity manifests as specific pressure points in the hardware. For instance, fine-grained meshes tend to stress cache hierarchies, while complex material behaviors increase compute-bound execution.
%
% All workloads are profiled using consistent thread affinity and fixed CPU frequency to ensure comparable results. Manual review of VTune traces is conducted to identify shared patterns, recurring bottlenecks, and function-level inefficiencies across the workload spectrum.

\vspace{-.6em}
\subsection{Gem5 Simulation Specifics}
% \Johnson{gem5 methodology and a table specifing the configurations and the range within which the parameters are being changed.}

\begin{table}[t]
\centering
\caption{Baseline CPU and system configuration in gem5.}
%\vspace{-5pt}
\begin{tabular}{ll}
\toprule
\textbf{Parameter} & \textbf{Value} \\
\midrule
ISA & x86 \\
Number of simulated cores & 2 \\
CPU model & X86O3CPU (OoO) \\
Core clock frequency & 3~GHz \\
Pipeline width (fetch/dispatch/issue/commit) & 4 / 6 / 6 / 4 \\
Rename width & 6 \\
Writeback / squash width & 8 / 6 \\
Reorder Buffer (ROB) entries & 224 \\
Issue Queue (IQ) entries & 128 \\
Load Queue / Store Queue entries & 72 / 56 \\
Integer / FP physical registers & 280 / 168 \\
L1I / L1D cache & 32~kB, 8-way \\
L2 cache & 1~MB, 16-way \\
MSHRs (L1I / L1D) & 32 / 32 \\
Cache line size & 64~B \\
Memory type & DDR4-2400 \\
Memory size & 4~GB \\
Branch predictor & TournamentBP \\
% BTB entries & 8192 \\
Operating system & Ubuntu 22.04.5  \\
\bottomrule
\end{tabular}
\label{tab:gem5-config}
\end{table}

% \Bahar{ready for review}

To explore architectural bottlenecks and evaluate microarchitectural sensitivities in biomechanics workloads, we use the gem5 simulator \cite{gem5} in full-system (FS) mode. All simulations are performed using the latest stable release at the time of this work (gem5 v24.0.0.1). We use the \texttt{X86O3CPU} core model, a detailed out-of-order CPU configured to represent a high-end Intel x86 microarchitecture, with support for wide dispatch, deep reordering, and aggressive speculative execution. The system runs a full Linux kernel (Ubuntu 22.04.5 with Linux 5.15.141) and boots from a custom filesystem image preloaded with the FEBio software stack and input models. This image is built using \texttt{packer} \cite{packer}, which automates the provisioning of root filesystems for gem5 full-system simulation.

Table~\ref{tab:gem5-config} summarizes the simulated microarchitectural baseline, including core pipeline parameters, cache hierarchy, and memory subsystem. We simulate two cores at 3GHz, each with private 32kB L1 instruction and data caches and a shared 1~MB L2 cache, backed by multi-channel DDR4-2400 DRAM. We restrict simulations to two cores due to the high computational cost of multi-core FS simulation in gem5. While FEBio employs OpenMP to exploit thread-level parallelism, such speedups mainly reflect throughput improvements rather than new architectural insights. Since our goal is to characterize core-level performance bottlenecks rather than parallel scaling behavior, a dual-core setup provides sufficient fidelity. To isolate application execution and reduce OS interference, we pin FEBio to one core and reserve the second for background system activity. To further reduce simulation time and focus on performance-critical phases, we launch each workload from a pre-created checkpoint immediately prior to solver execution. While we do not run VTune within gem5, our experimental design is guided by VTune profiles from real hardware, which consistently highlight architectural bottlenecks such as time spent in spin-wait loops, instruction serialization from data dependencies, and backend stalls caused by irregular memory access and poor cache reuse.

\section{Results \& Analysis}

 In this section, we present a detailed analysis of biomechanics workloads using a combination of real-system profiling and architectural simulation. We first use Intel VTune \cite{intel2023vtune} to identify key performance bottlenecks—such as memory stalls, control flow inefficiencies, and backend pressure, across representative FEBio models. Guided by these observations, we then transition to full-system gem5 \cite{gem5} simulations to conduct controlled sensitivity studies. By sweeping parameters such as cache sizes, pipeline widths, and queue depths, we assess how architectural choices influence execution efficiency and identify opportunities for hardware specialization in domain-specific accelerators. 

%comment 1 was here

\begin{figure}[t]
    \centering
    \includegraphics[width=\columnwidth]{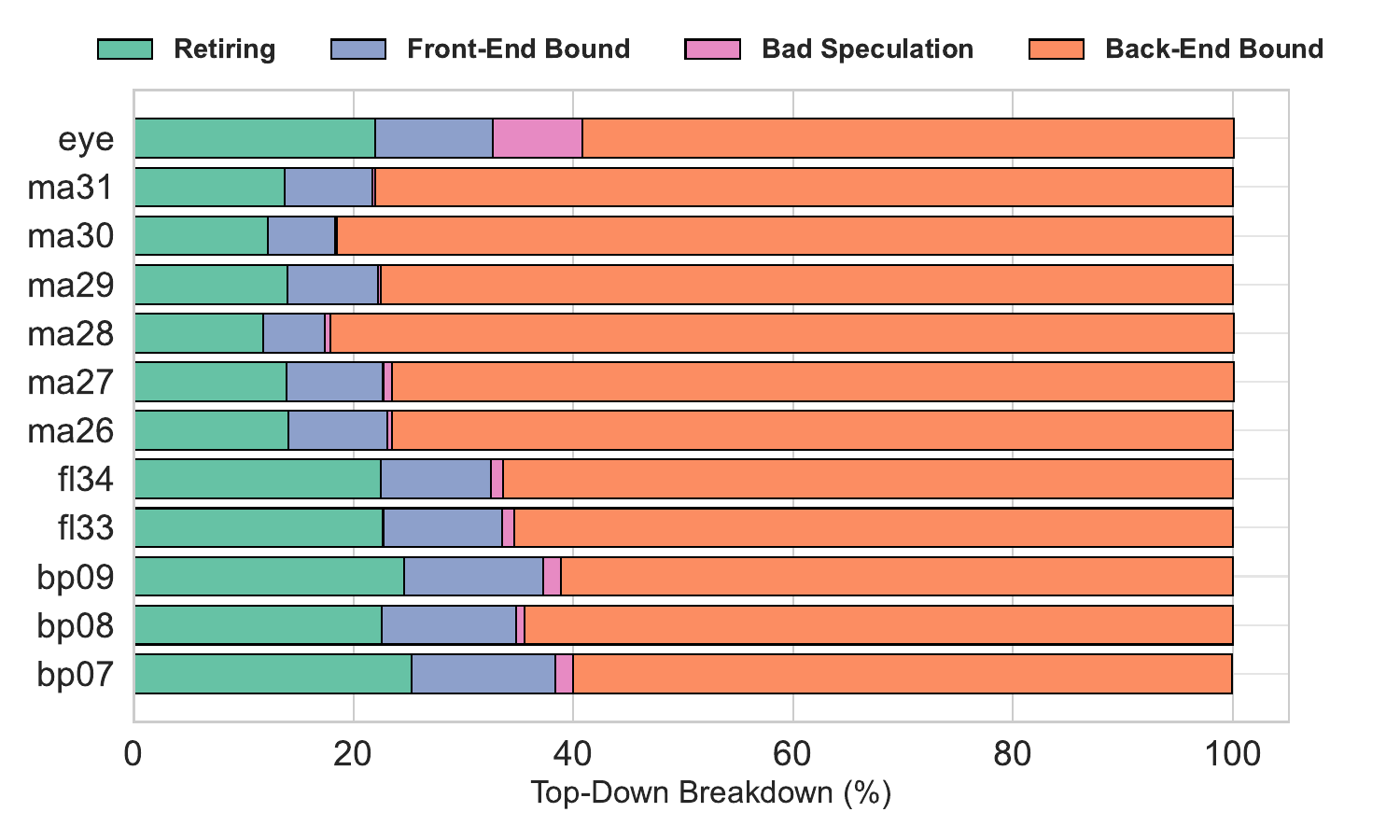}
    \vspace{-15pt}
    \caption{Top-down pipeline breakdown.}
    \vspace{-10pt}
    \label{fig:topdown_stacked}
\end{figure}

%comment 2 was here
\begin{figure}[b]
    \centering
    \includegraphics[width=\linewidth]{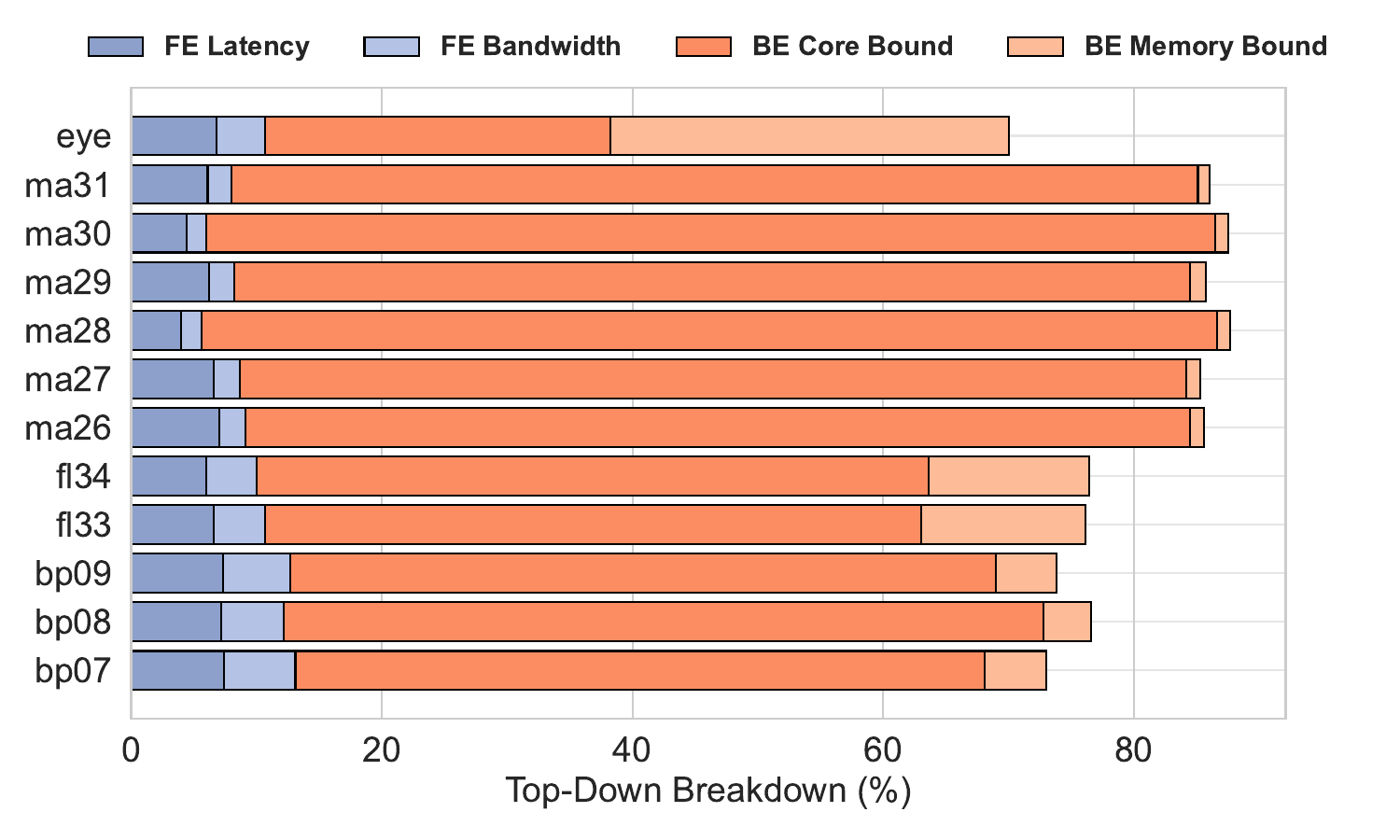}
    \caption{Breakdown of front-end (FE) and back-end (BE) pipeline stalls across FEBio workloads. Bad speculation is excluded from the plots as its contribution is negligible.}
    \vspace{-12pt}
    \label{fig:fe_be_breakdown}
\end{figure}

\subsection{Architectural Bottlenecks in FEBio}

% \Amirmahdi{updating section with consistent workloads}

% \begin{figure}[t]
%     \centering
%     \includegraphics[width=\columnwidth]{figs/topdown_stacked_horizontal_seabornstyle.pdf}
%     \caption{Top-down pipeline breakdown for FEBio workloads.}
%     \label{fig:topdown_stacked}
% \end{figure}

We use top-down microarchitectural analysis to understand how execution time is distributed across the pipeline for each workload. This methodology divides pipeline slots into four main categories: Retiring, Front-End Bound, Bad Speculation, and Back-End Bound. The breakdown for each model is presented in Figure~\ref{fig:topdown_stacked}.

\begin{figure*}[h]
\centering
  \vspace{-10pt}
  \includegraphics[width=0.9\textwidth]{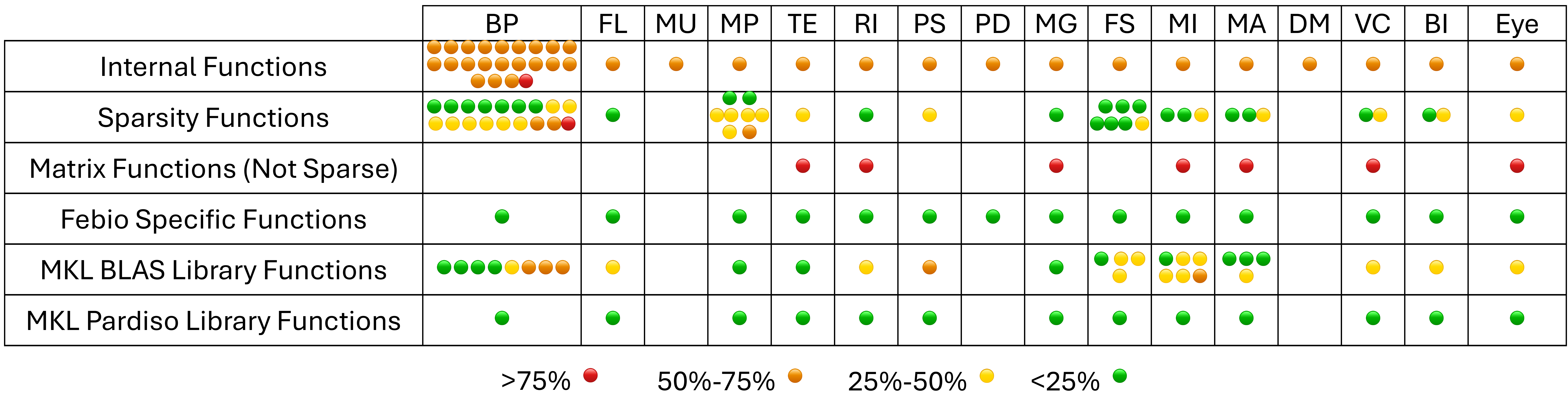}
  \caption{Prevalence of common function types within the top 5\% of clockticks across workloads. Dot color reflects the fraction of top hotspots from each function category.}
  % : red (>75\%), orange (50–75\%), yellow (25–50\%), and green (<25\%).}
  \vspace{-10pt}
 
  \label{fig:hotspots}
\end{figure*}

Most workloads exhibit clear pipeline inefficiencies, as summarized in Figure \ref{fig:topdown_stacked}. Biphasic (\texttt{bp07}, \texttt{bp08}, \texttt{bp09}) and fluid (\texttt{fl33}, \texttt{fl34}) models demonstrate modest instruction retirement rates around $22 -25\%$, indicating considerable pipeline stalls primarily driven by backend constraints. Specifically, these models show significant backend-bound percentages (ranging from approximately $59.9\%$ to $66.4\%$), highlighting intense memory latency and core resource limitations characteristic of computationally demanding models. Material models (\texttt{ma26–ma31}) exhibit even lower instruction retirement rates at $11.8-14.1\%$, underscoring extreme backend pressure. Models such as \texttt{ma28} and \texttt{ma30} show backend-bound rates exceeding $80\%$, reflecting severe contention for memory and execution units. Front-end bound stalls are uniformly moderate at $5.6-13.1\%$ across all workloads, suggesting only modest bottlenecks in instruction fetch or decoding phases. Bad speculation rates remain minimal (mostly below $1\%$) across all models, indicative of relatively straightforward computational kernels with infrequent branching, particularly pronounced in material models. Collectively, these results highlight that FEBio workloads predominantly experience severe backend constraints, especially in simpler material models where low retirement rates reflect pronounced memory hierarchy and execution queue saturation. Consequently, optimizations targeting memory subsystem improvements and increased backend resources would significantly enhance performance across these models.

% %comment 2 was here
% \begin{figure}[h]
%     \centering
%     \includegraphics[width=\linewidth]{figs/new_fe_be.pdf}
%     \caption{Breakdown of front-end (FE) and back-end (BE) pipeline stalls across FEBio workloads. Bad speculation is excluded from the plots as its contribution is negligible.}
%     \vspace{-12pt}
%     \label{fig:fe_be_breakdown}
% \end{figure}

% \begin{figure*}[t]
% \centering
%   \vspace{-10pt}
%   \includegraphics[width=0.9\textwidth]{figs/Common_Functions.pdf}
%   \caption{Prevalence of common function types within the top 5\% of clockticks across workloads. Dot color reflects the fraction of top hotspots from each function category.}
%   % : red (>75\%), orange (50–75\%), yellow (25–50\%), and green (<25\%).}
%   \vspace{-10pt}
 
%   \label{fig:hotspots}
% \end{figure*}

Figure~\ref{fig:fe_be_breakdown} expands on the top-level stall categories depicted earlier, providing detailed insights into the front-end and back-end bound cycles. Front-end latency accounts for stalls from instruction cache misses, whereas front-end bandwidth limitations occur when fewer instructions are fetched than the theoretical maximum. Backend stalls are categorized into memory-bound (cycles waiting on unresolved loads or cache misses) and core-bound stalls (cycles constrained by limited execution resources or execution unit contention). Biphasic (\texttt{bp07}, \texttt{bp08}, \texttt{bp09}) and fluid models (\texttt{fl33}, \texttt{fl34}) demonstrate significant memory-bound stalls, with fluid models particularly exhibiting high memory pressure (around $13\%$), emphasizing their dependency on memory hierarchy performance. Material models (\texttt{ma26–ma31}) are overwhelmingly core-bound ($75-81\%$), mostly because of serialization pressure from repeated slow PAUSE instructions \cite{cloutier_pause}, highlighting severe limitations in execution resources rather than memory. Front-end latency and bandwidth constraints are comparatively minor across all models (typically below $8\%$), indicating that optimization strategies should primarily focus on addressing backend core-resource bottlenecks for material models and memory-hierarchy improvements for fluid and biphasic models.

 Across the FEBio test suite and the ocular case study, we observe that the models placing the greatest stress on the system share three key biomechanical characteristics: (1) high element and node counts, (2) nonlinear or multiphasic material models, and (3) complex boundary conditions such as time-dependent loading or contact mechanics. Models featuring biphasic or multiphasic materials consistently incur longer runtimes and higher memory-bound stall percentages. These material formulations introduce additional field variables (e.g., pore pressure, solute concentration), increasing both the size and sparsity of the global stiffness matrix. As a result, matrix assembly becomes more irregular, and memory access patterns lose spatial locality. Similarly, simulations involving large deformations or contact enforcement (e.g., joint mechanics models) show elevated cache miss rates and deeper memory hierarchy dependencies, reflecting the nonlinear, data-dependent nature of these problems. The ocular biomechanics model,
 %with over 150,000 degrees of freedom, nonlinear tissue behavior, 
 produces the most sustained pressure on backend execution resources. In this model, VTune reports frequent L2 and LLC cache misses, high memory-bound stall cycles, and significant pipeline underutilization. These results confirm that physiologically realistic, fine-grained models exhibit architectural behavior that is substantially more demanding than synthetic test cases, as a result of their sparsity, irregularity, and iteration-intensive solvers.

\subsubsection{Function Level Profiling}
\label{sec:hotspot}
%i agree that it should be left in 
% \Jnote{Though commented out, Hana thinks section this is not important enough. However, I think it'd be good to have if we have space. Its figure gives insights into hotspot functions and consequently areas of DSA optimizations. Remove or keep? \Hnote{keep}}

\begin{figure}[b]
\centering
  \includegraphics[width=0.9\columnwidth]{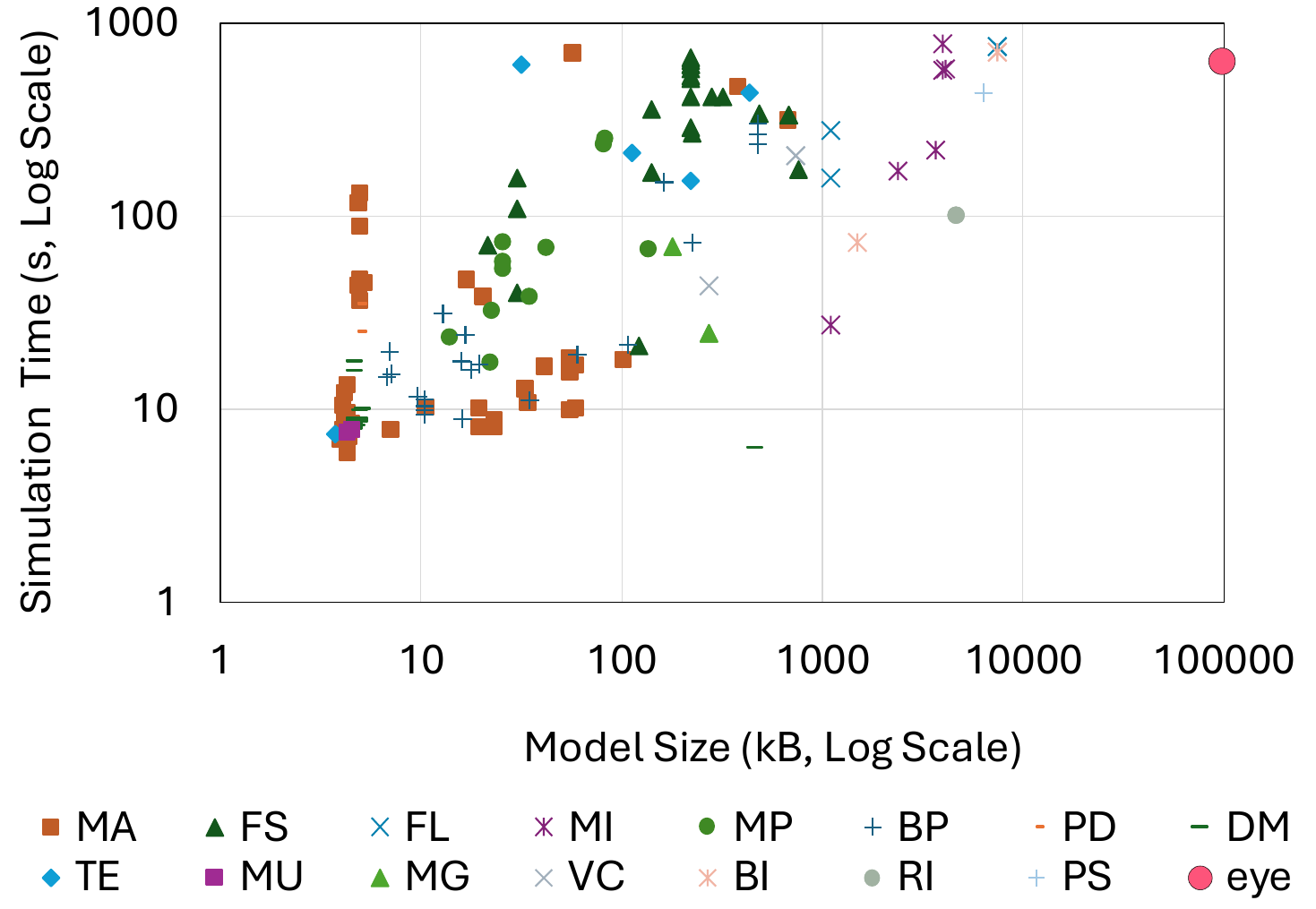}
  \caption{Simulation time and model size for the \texttt{eye} model and test suite models. 
   % The line of best fit is updated and is used to identify any outliers in the data. 
   The models are organized by size and mapped onto a logarithmic scale.
      }
      \vspace{-5pt}
   \label{fig:eyeSize}
 \end{figure}

To better understand where architectural bottlenecks manifest at the function level, we conducted a hotspot analysis using VTune’s bottom-up reporting. Figure~\ref{fig:hotspots} summarizes the dominant functions, those contributing to the top 5\% of clock ticks, across representative FEBio workloads. Functions are grouped into six categories as shown in the illustration. Each cell uses color-coded dots to show the prevalence of these function types within the top 5\% of hotspots. The number of dots in each cell represents how many functions from that category meet the threshold. If all contributions come from a single function, we simplify the display by using just one dot.

%The number of dots in each cell represents the number of functions from that function type within the category meeting the threshold, simplified as far as possible. For instance, the four green dots under BP’s MKL BLAS Library Functions indicate that 50\% of the top 5\% of clockticks were seen in BP functions, but that each MKL BLAS Library routine contributed less than 25\% of the total clock ticks.

\begin{center}
\begin{figure}[b]
\vspace{-10pt}
   \centering
    \includegraphics[width=0.9\columnwidth]{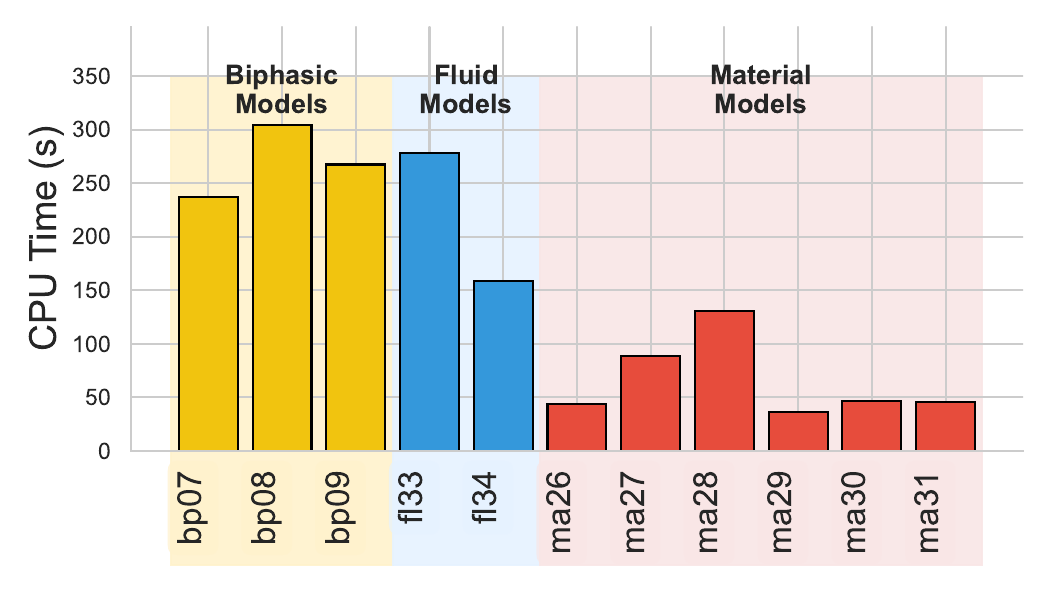}
  \vspace{-10pt}
  \caption{Execution time of models visualized to see the performance discrepancies in CPU time.
  % \Amirmahdi{Please can you reduce the height by 1/3 to save space.}
  % \Jnote{This plot appears to include a different set of models than the 11 selected by Professor Babak for gem5 experiments. Would it be possible to focus on that same set for consistency? Specifically, do you think it's necessary to have data for bp and fl replace pd and dm? }
  }
  \label{fig:zoomedSize}
\end{figure}
\end{center}

\vspace{-3em}

Across nearly all workloads, \textbf{internal functions} dominate the profile. For example, BP workload shows red and orange indicators in this row, revealing that between 50–100\% of its top hotspots come from FEBio’s stiffness matrix assembly, residual computation, and force evaluation routines. These operations align with earlier memory-bound characterization, exhibiting sparse, element-wise memory access and limited vectorization. \textbf{Sparsity functions} also appears frequently but are less dominant, often present with green dots, indicating a modest contribution per function. Workloads such as BP and MI also show increased reliance on \textbf{MKL BLAS routines} and matrix operations, reflecting denser numerical kernels. The \texttt{eye}, a high-resolution case study, displays dispersed hotspots across all categories, consistent with deeper call graphs and more diverse execution paths. These insights emphasize that FEBio's performance bottlenecks are rooted not in highly regular computation, but in sparse, memory- and branch-heavy functions. Future DSAs should prioritize accelerating these irregular paths over dense linear algebra acceleration.

% \begin{tcolorbox}[enhanced,width=3.4in,
%     drop fuzzy shadow southeast,
%     colframe=gray!10,colback=gray!10]
% \textbf{Takeaway II:} FEBio performance bottlenecks stem primarily from irregular, memory- and branch-intensive functions such as stiffness assembly and residual evaluation. Accelerators targeting FEBio workloads should prioritize sparse, data-dependent paths rather than focusing solely on dense compute kernels.
% \end{tcolorbox}

% \begin{figure}[!t]
% \centering
%   \includegraphics[width=0.9\columnwidth]{figs/Execution_time_size.pdf}
%   \caption{Simulation time and model size for the eye model and test suite models. 
%    % The line of best fit is updated and is used to identify any outliers in the data. 
%    The models are organized by size and mapped onto a logarithmic scale. 
%    % \Jnote{I don't see a line of best fit in the plot.} 
% %   %  \Hana{Bahar: I couldn't find the editable source of the single-color chart, with the eye model label on it. I've added the edited version of this color-coded chart instead (in pptx format in source files). Please update it with marking marking the eye model. }
% %   % \Hana{Make sure all the model names used in this chart are defined in the table in workload composition section.}
%       }
%       \vspace{-5pt}
%    \label{fig:eyeSize}
%  \end{figure}

\setlength{\textfloatsep}{8pt}

\subsubsection{Workload Scaling Analysis}
\label{sec:workload scaling}
In addition to microarchitectural profiling, we examine how simulation performance scales with model size and complexity. As a surrogate for model complexity we use the test models' input file size, which correlates with the number of nodes and elements of the model (mesh size). Although "complexity" is not easy to define objectively, the mesh size and by extension the size of the input file is an appropriate  practical measure. As shown in Figure~\ref{fig:eyeSize}, simulation time generally increases with model size, but the relationship is not strictly linear. Most test suite models follow a sublinear or near power-law trend, indicating that computational cost scales reasonably with increasing problem size. The case study model, \texttt{eye}, however, lies well above this curve, reflecting disproportionately higher execution time relative to similarly sized workloads. This deviation suggests that as models grow in size, mesh complexity, and stiffness matrix density, architectural overheads, such as cache miss rates, memory latency, and serialization, become dominant. These effects are compounded by irregular memory access patterns and deeper reuse distances, which expose back-end limitations and lead to poor memory-level parallelism. Consequently, performance degrades more rapidly than size alone would suggest, reinforcing earlier observations that large biomechanics models are predominantly back-end and memory bound.

Performance variation is not solely determined by the initial characteristics of a model (i.e. size). As illustrated in Figure \ref{fig:zoomedSize}, models with similar sizes, such as biphasic, fluid, or material, can exhibit significantly different execution times. Particularly, biphasic and fluid models generally require substantially more computational resources, reflected by their longer execution times compared to material models. This variation highlights the need to look at more detailed model parameters (i.e. element count, mesh density, etc.) and the computational intensity introduced by the complex physics involved in biphasic and fluid models, demanding extensive constitutive updates, increased nonlinear solver iterations, and prolonged matrix assembly phases. Consequently, these additional computational requirements amplify backend and memory pressure, independent of the general category to which the model belongs.

% \begin{tcolorbox}[enhanced,width=3.4in,
%     drop fuzzy shadow southeast,
%     colframe=gray!10,colback=gray!10]
% \textbf{Takeaway III:} Scaling behavior varies across workloads. While smaller test models follow a sublinear trend, larger case studies such as \texttt{eye\_model} introduce disproportionately higher memory and execution pressure, resulting in steeper performance degradation. This highlights the need for deeper memory hierarchies and backend resources in DSAs targeting large-scale biomechanical simulations.
% \end{tcolorbox}

\begin{tcolorbox}[enhanced,width=3.4in,
    drop fuzzy shadow southeast,
    colframe=gray!10,colback=gray!10]
\textbf{Takeaway I:}
FEBio’s performance bottlenecks arise mainly from irregular memory access and sparse computations, such as stiffness matrix assembly and residual evaluation.
%, rather than dense kernels. 
These bottlenecks increase
%become more pronounced 
with higher mesh density, complex boundary conditions, and larger workloads. 
%While small models scale sublinearly, [larger]
Large cases like \textit{eye} see disproportionately higher memory and execution demands, highlighting FEBio’s sensitivity to workload characteristics and the growing impact of memory behavior in large-scale simulations.
%FEBio performance bottlenecks arise primarily from irregular, memory accessing and sparse computation operations such as stiffness matrix assembly and residual evaluation, rather than from dense numerical kernels. These bottlenecks become more pronounced with increased mesh density, boundary constraints, and workload size. While smaller test models show sublinear scaling, larger case studies like \textit{eye} experience disproportionately higher memory and execution pressure, resulting in steeper performance degradation. This variation highlights the sensitivity of FEBio’s performance to workload characteristics and the growing impact of memory behavior in large-scale biomechanical simulations.
\end{tcolorbox}

\begin{figure*}
    \centering
    \begin{subfigure}[t]{0.32\textwidth}
        \centering
        \includegraphics[width=\linewidth]{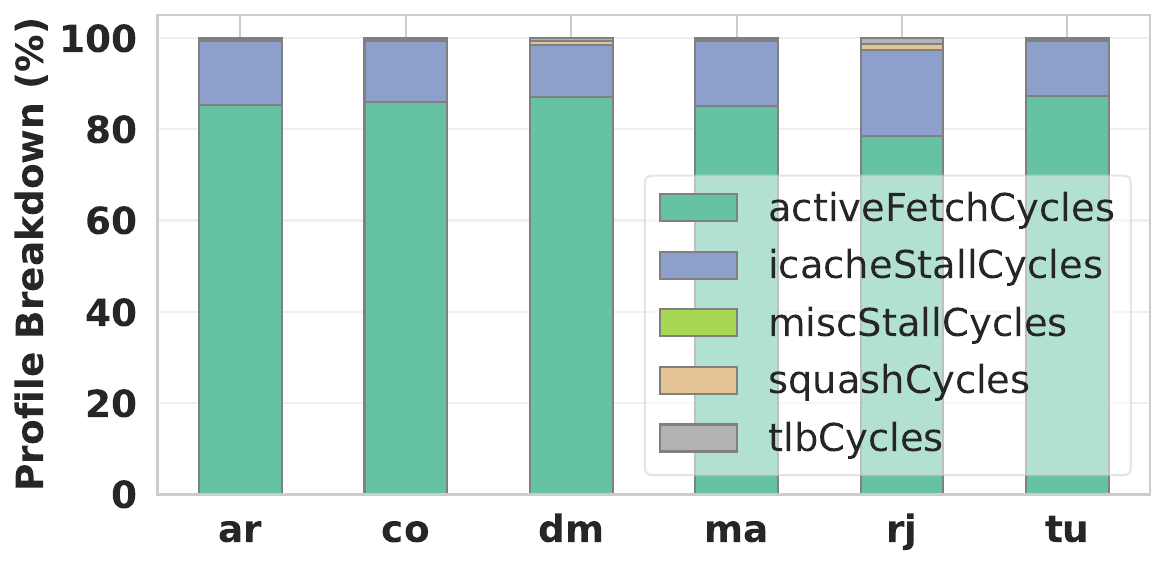}
        \caption{Fetch stage characteristics}
        \label{fig:char_fetch}
    \end{subfigure}
    \hfill
    \begin{subfigure}[t]{0.32\textwidth}
        \centering
        \includegraphics[width=\linewidth]{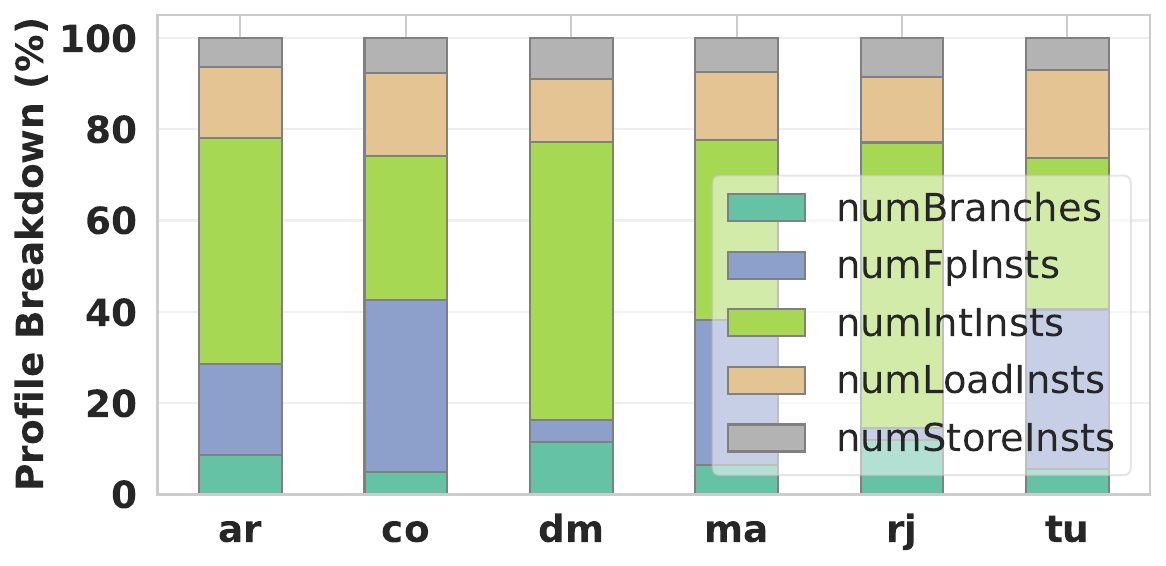}
        \caption{Execute stage characteristics}
        \label{fig:char_execute}
    \end{subfigure}
    \hfill
    \begin{subfigure}[t]{0.32\textwidth}
        \centering
        \includegraphics[width=\linewidth]{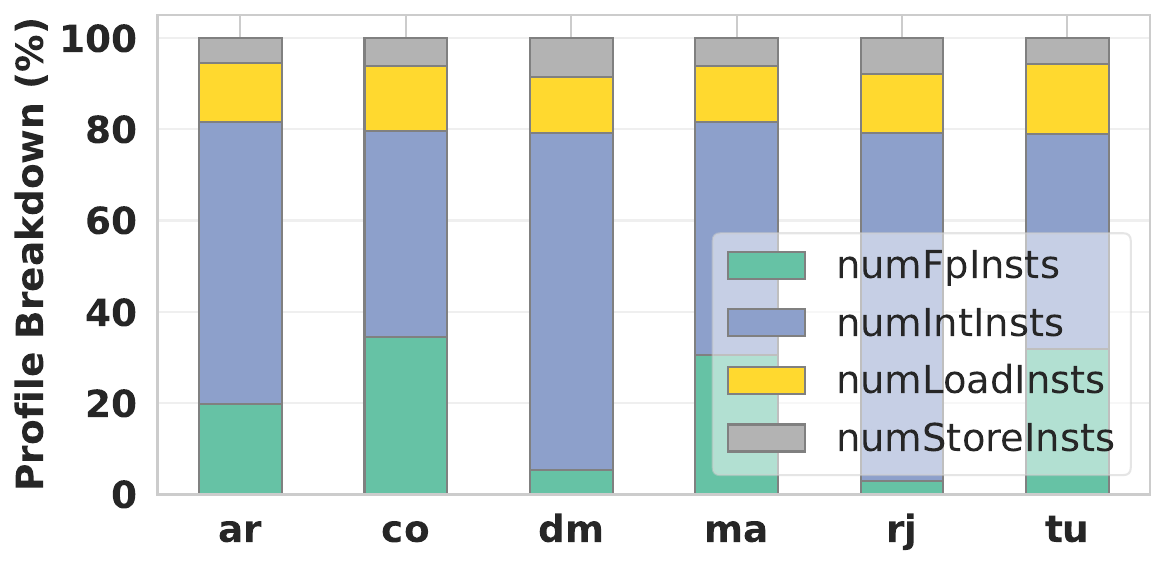}
        \caption{Commit stage characteristics}
        \label{fig:char_commit}
    \end{subfigure}
    \caption{Normalized pipeline-stage activity breakdown for FEBio workloads across the fetch, execute, and commit stages.}
    \vspace{-10pt}
    \label{fig:pipeline_characterization}
\end{figure*}

% \subsection{Characterizing FEBio Models}
% \Jnote{I suggest this should not be a stand-alone section. It can be summarized and added as a conclusion to the section on top-down analysis}

% \begin{tcolorbox}[enhanced,width=3.4in,
%     drop fuzzy shadow southeast,
%     colframe=gray!10,colback=gray!10]
% \textbf{Takeaway IV:} Biomechanical workloads that combine large model sizes, nonlinear or multiphasic materials, and complex boundary conditions place the greatest stress on modern architectures. These models incur higher memory-bound stalls, cache miss rates, and pipeline inefficiencies due to increased sparsity, irregular memory access patterns, and solver complexity—highlighting the performance demands of physiologically realistic simulations. 

% \end{tcolorbox}
%\vspace*{-.75cm}
\subsection{Instruction-Level Pipeline Characterization}
Before evaluating microarchitectural sensitivities, we characterize how each FEBio workload stresses the core pipeline under the baseline gem5 configuration. The \texttt{X86O3CPU} model in gem5 implements a detailed out-of-order pipeline based on the Alpha 21264, with realistic modeling of fetch, execution, and commit stages. Analyzing instruction-level activity across these stages reveals workload-specific bottlenecks that guide our hardware parameter sweeps.

Figure~\ref{fig:char_fetch} shows the breakdown of fetch-stage activity. Here, \texttt{cycles} represents time spent actively fetching instructions, while \texttt{icache stall cycles} and \texttt{TLB cycles} capture stalls as a result of instruction cache misses and address translation, respectively. Other components like miscellaneous stall or squash cycles indicate backend-induced or system-level delays such as trap handling or pipeline flushes. All workloads exhibit a significant contribution of 78--87\% from regular fetch cycles, but models such as \texttt{rj} show elevated instruction cache stall cycles, suggesting higher frontend pressure. 
% The \texttt{tu} model exhibits the most balanced fetch profile, with modest stalls and fewer squashes.
Figure~\ref{fig:char_execute} depicts the execution stage activity breakdown. \texttt{co} stands out with the highest proportion of memory operations, with load/store instructions accounting for $\sim$26\% of execution-stage activity, consistent with its irregular memory access patterns. In contrast, \texttt{ar} and \texttt{ma} workloads show a heavier skew toward integer and floating-point instructions, reflecting more regular numerical kernels. Figure~\ref{fig:char_commit} illustrates the commit stage characterization. \texttt{dm} and \texttt{rj} commit stages are moderately dominated by memory references, with lower floating-point activity, similar to the execute stage observation. Meanwhile, \texttt{tu}, \texttt{ma}, and \texttt{co}, mirrowing the execute stage, feature a higher share of floating-point instructions, ranging from 31--37\%, consistent with their more core-bound material models.

These differences underscore the architectural diversity of FEBio workloads. While some models are frontend- or core-bound, others present back-end or memory-centric stress. While some have disproportionate integer and floating-point instructions, others are more balanced. This motivates our targeted exploration of core pipeline widths, queue depths, and memory subsystem configurations in the following sections.

% \begin{tcolorbox}[enhanced,width=3.4in,
%     drop fuzzy shadow southeast,
%     colframe=gray!10,colback=gray!10]
% \textbf{Takeaway V:} FEBio workloads exhibit heterogeneous instruction profiles across pipeline stages. Most time is spent in regular fetch cycles, but frontend stalls (e.g., icache misses) are non-negligible in some models. Execution and commit stages are dominated by integer and floating-point operations, with memory instructions playing a secondary role.
% \end{tcolorbox}

\vspace*{-.2cm}
\subsection{gem5 Microarchitectural Sensitivity Studies}

To evaluate how biomechanics workloads respond to microarchitectural variation, we conduct a series of sensitivity studies in gem5. Each experiment isolates a specific hardware parameter, such as pipeline width, buffer capacity, or cache size, while holding others fixed. These studies are guided by VTune profiling, which reveals that FEBio workloads could be front-end or backend-bound, memory-latency-sensitive, and exhibit limited instruction-level parallelism (ILP) because of sparse computation and synchronization, depending on the workload. 
% By analyzing performance trends across these sweeps, we identify which architectural resources most strongly influence execution efficiency and where targeted specialization offers the greatest benefit.

\begin{figure}
    \centering
    \begin{subfigure}[t]{0.24\textwidth}
        \centering
        \includegraphics[width=\linewidth]{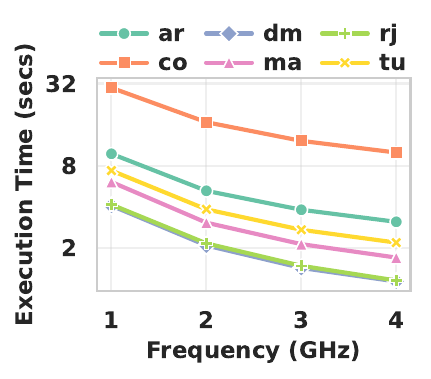}
        \caption{Execution time vs. frequency.}
        \label{fig:freq_sweep_time}
    \end{subfigure}%
    \hfill
    \begin{subfigure}[t]{0.24\textwidth}
        \centering
        \includegraphics[width=\linewidth]{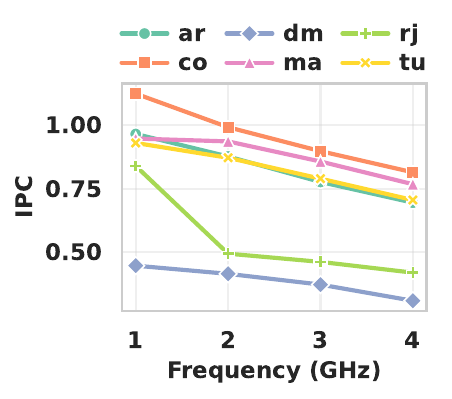}
        \caption{IPC vs. frequency.}
        \label{fig:freq_sweep_ipc}
    \end{subfigure}
    \caption{Core frequency sensitivity across FEBio workloads.}
    \vspace{-5pt}
    \label{fig:freq_sweep}
\end{figure}

\vspace*{-.1cm}
\subsubsection{Sensitivity to Core Frequency}
% \Johnson{gem5 results}

We vary core frequency to assess the extent to which FEBio workloads are limited by core throughput versus memory or synchronization latency, and observe changes in performance. This experiment is motivated by the need to evaluate whether frequency scaling yields proportional speedup, which is key to determine whether a domain-specific accelerator should invest in deeper pipelines or higher frequency operation.

Figure~\ref{fig:freq_sweep_time} shows that execution time decreases sub-linearly with increasing frequency, indicating that workloads are not fully compute-bound. For example, with respect to a 1\,GHz clock, \texttt{ma} and \texttt{dm} see 2.86$\times$ speedup at 3\,GHz, while \texttt{co} achieves only 2.45$\times$, despite a 3$\times$ increase in clock rate. Even at 4\,GHz, most models fall short of ideal scaling: \texttt{ar} achieves 3.15$\times$ speedup and \texttt{co} 2.98$\times$, revealing diminishing returns due to backend stalls and memory access delays. Figure~\ref{fig:freq_sweep_ipc} supports this conclusion: IPC tends to decrease as frequency increases, indicating that faster clocks expose backend and memory-related stalls more prominently. As the processor attempts to retire instructions more quickly, unresolved dependencies and memory latency become increasingly dominant, limiting instruction throughput. This effect is especially visible in \texttt{dm} and \texttt{rj}, where IPC drops by over 30\% from 1\,GHz to 4\,GHz. These diminishing returns align with our fetch-stage characterization (Figure~\ref{fig:char_fetch}), where \texttt{rj} exhibit relatively higher icache and trap-induced stalls. Such front-end inefficiencies limit throughput scaling at higher clock rates.

% \begin{figure}[h]
%     \centering
%     \begin{subfigure}[t]{\linewidth}
%         \centering
%         \includegraphics[width=\linewidth]{figs/l1i_mpki_vs_cache_size.pdf}
%         \caption{L1 instruction cache miss rate.}
%         \label{fig:l1i_mpki_vs_cache_size}
%     \end{subfigure}
    
%     \begin{subfigure}[t]{\linewidth}
%         \centering
%         \includegraphics[width=\linewidth]{figs/l1d_mpki_vs_cache_size.pdf}
%         \caption{L1 data cache miss rate.}
%         \label{fig:l1d_mpki_vs_cache_size}
%     \end{subfigure}
    
%     \begin{subfigure}[t]{\linewidth}
%         \centering
%         \includegraphics[width=\linewidth]{figs/l1d_mpki_vs_cache_size.pdf}
%         \caption{L2 cache miss rate.}
%         \label{fig:l1d_mpki_vs_cache_size}
%     \end{subfigure}
    
%     \caption{Miss rate sensitivity to cache capacity across FEBio workloads.}
%     \label{fig:cache_missrate_sensitivity}
% \end{figure}

\begin{figure*}[h]
    \centering
    \begin{subfigure}[t]{0.19\textwidth}
        \centering
        \includegraphics[width=\linewidth]{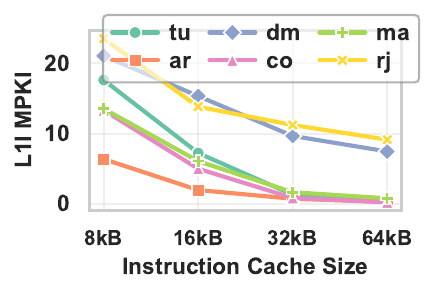}
        \caption{L1I MPKI}
        \label{fig:l1i_mpki_vs_cache_size}
    \end{subfigure}
    \hfill
    \begin{subfigure}[t]{0.19\textwidth}
        \centering
        \includegraphics[width=\linewidth]{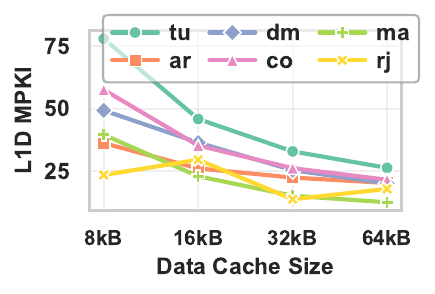}
        \caption{L1D MPKI}
        \label{fig:l1d_mpki_vs_cache_size}
    \end{subfigure}
    \hfill
    \begin{subfigure}[t]{0.19\textwidth}
        \centering
        \includegraphics[width=\linewidth]{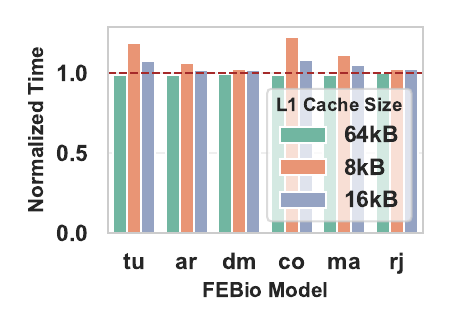}
        \caption{L1 exec time (norm.)}
        \label{fig:l1_exec_time}
    \end{subfigure}
    \hfill
    \begin{subfigure}[t]{0.19\textwidth}
        \centering
        \includegraphics[width=\linewidth]{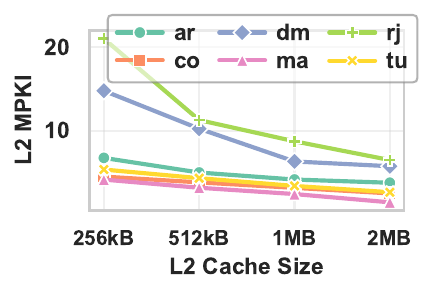}
        \caption{L2 MPKI}
        \label{fig:l2_mpki_vs_cache_size}
    \end{subfigure}
    \hfill
    \begin{subfigure}[t]{0.19\textwidth}
        \centering
        \includegraphics[width=\linewidth]{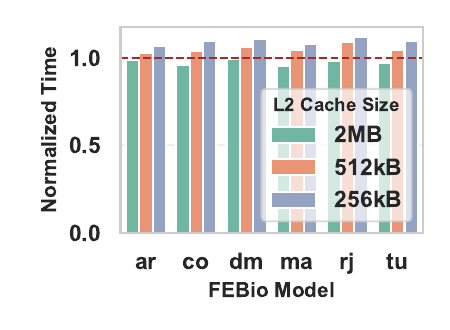}
        \caption{L2 exec time (norm.)}
        \label{fig:l2_exec_time}
    \end{subfigure}
    \caption{Cache sensitivity analysis across FEBio workloads. MPKI (a, b, d) and normalized execution time (c, e) are shown for varying L1 and L2 cache sizes.}
    \vspace{-10pt}
    \label{fig:cache_sensitivity_all}
\end{figure*}

%\vspace*{-.3cm}
\subsubsection{Sensitivity to Cache Capacity}

FEBio workloads exhibit varied data locality behaviors across models. To understand the role of temporal locality and working set size in biomechanics simulation, we sweep the sizes of both L1 and L2 caches.
% while holding all other microarchitectural parameters fixed 
% This experiment is motivated by VTune profiling, which shows that while many smaller models are backend-bound with low memory pressure, others can exhibit pronounced memory-bound behavior. These effects stem from factors like mesh resolution and stiffness matrix size, which directly impact memory access intensity and solver behavior. 
Our goal is to quantify how cache capacity influences Misses Per Kilo Instruction (MPKI), and eventually execution time, enabling us to identify minimum effective sizes that preserve performance. These insights are particularly relevant for cache provisioning in area- and power-constrained on-chip or off-chip accelerators like those for scientific computing.

\paragraph{\textbf{L1 Cache Sensitivity}}
Figures~\ref{fig:l1i_mpki_vs_cache_size} and~\ref{fig:l1d_mpki_vs_cache_size} show L1I and L1D MPKI trends for cache sizes ranging from 8\,kB to 64\,kB. Instruction cache behavior converges quickly: all workloads exhibit sharp drops from 8\,kB to 32\,kB, after which MPKI largely stabilizes. For example, \texttt{tu} shows a 16.34 absolute drop in L1I MPKI between 8\,kB and 32\,kB, with very minimal improvement at 64\,kB. \texttt{rj} followed by \texttt{dm}, show the highest sensitivities to the L1I cache, hinting at their large instruction footprint or poor data locality. \texttt{ar}, on the other hand, shows the least sensitivity to L1I cache. L1D trends are more diverse. The drops are especially large from 8\,kB to 16\,kB, then gradually taper off. Memory-sensitive models such as \texttt{co} and \texttt{tu} experience large gains—L1D MPKI falls from 57.6 and 77.9 at 8\,kB to 26.2 and 32.9 at 32\,kB, respectively. Meanwhile, \texttt{rj} exhibits relatively low L1D MPKI, following a non-monotonic pattern that rises and falls, indicating cache thrashing or an irregular memory access pattern, typical of sparse matrix workloads. Execution time trends in Figure~\ref{fig:l1_exec_time} confirm that a 32\,kB L1 cache is a practical inflection point, most workloads achieving their best or near-best performance. For example, \texttt{co}'s execution time is 1.23$\times$ faster at 32\,kB than at 8\,kB, with very minimal benefit at 64\,kB. Likewise, \texttt{tu} execution time improves by 1.18$\times$. Moreover, \texttt{rj}'s execution time declines slightly beyond 32\,kB. These diminishing returns suggest that increasing beyond 32\,kB offers little performance boost, due to the application's working set size and longer access latency associated with larger caches. 
% Thus, a 32\,kB instruction and data L1 cache strikes an effective balance between area, power, and locality capture for biomechanical workloads. Accelerator designs can safely avoid overprovisioning L1 caches/scratchpads without incurring meaningful penalties.

\paragraph{\textbf{L2 Cache Sensitivity}}
Figures~\ref{fig:l2_mpki_vs_cache_size} and~\ref{fig:l2_exec_time} show the impact of L2 cache capacity on MPKI and execution time, respectively. As illustrated, \texttt{rj} and \texttt{dm} exhibit notable sensitivity to the L2 cache, with the former flattening out at 512\,kB and the latter at 1\,MB. This suggests that \texttt{rj} with non-monotic or irregular behavior at L1 data cache benefits from larger, more associative L2 caches.
% For instance, between 256\,kB and 1\,MB, \texttt{ar} sees its L2 MPKI drop from 64.3\% to 38.8\%, and \texttt{dm} sees a drop from 48.2\% to 20.9\%. 
% Beyond 1\,MB, MPKI reductions are typically less prominent. 
Even though L2 size scaling is still effective up to 2MB, some models like \texttt{ar}, \texttt{ma}, \texttt{co}, and \texttt{tu} show L2 cache efficiency with less than 1\% misses in each thousand instructions across all evaluated cache sizes. Execution time trends assert this observation. \texttt{co}, \texttt{ma}, and \texttt{tu}, for instance, continue to show some gain at 2\,MB. Conversely, \texttt{dm} and \texttt{ar} exhibit limited improvement beyond 1\,MB. This sensitivity reflects deeper reuse distances or less predictable access patterns, indicating that cache hierarchy design can be workload-aware. 

The cache-related performance trends are also reflected in the execution and commit profiles (Figures~\ref{fig:char_execute},~\ref{fig:char_commit}), where relatively memory-bound workloads such as \texttt{co} and \texttt{tu} exhibit high proportions of loads and stores, making them more sensitive to L1 and L2 capacity changes.

% These observations support a tiered provisioning strategy: workloads with compact data working sets (e.g., \texttt{co}, \texttt{ma}) are well served by a 512\,kB L2, while those with irregular or sparse memory access (e.g., \texttt{dm}) benefit from up to 1–2\,MB. Importantly, all models demonstrate diminishing returns beyond 1\,MB, making it a practical upper bound for most DSA cache hierarchies. This enables resource reallocation toward memory-level parallelism or backend width without compromising performance.

\begin{figure}[t]
    \centering
    \includegraphics[width=\linewidth]{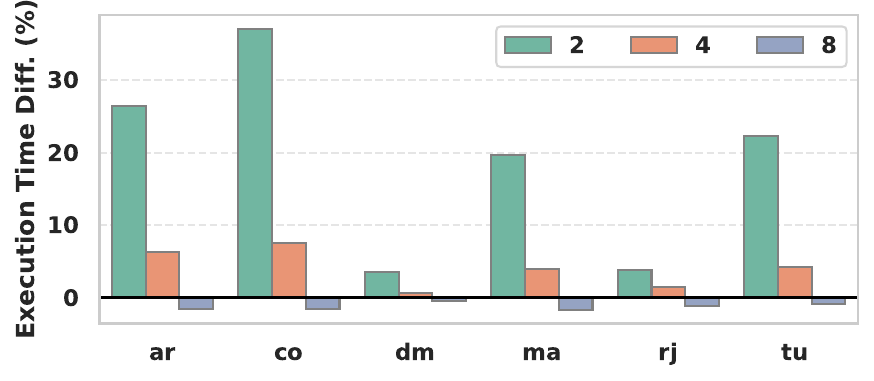}
    \caption{Execution time difference relative to baseline pipeline width=6.}
    \vspace{-10pt}
    \label{fig:width_sensitivity}
\end{figure}

\subsubsection{Sensitivity to Pipeline Widths}

Modern out-of-order processors rely on wide pipelines to exploit ILP, but the effectiveness of width scaling depends on workload characteristics. To assess the impact of pipeline parallelism on FEBio workloads, we vary core pipeline widths, specifically fetch, decode, rename, dispatch, and issue widths, in our gem5 model. This experiment targets ILP exposure: increasing width allows more instructions to be fetched and executed per cycle, which can improve throughput if workloads contain sufficient ILP. However, widening these stages may offer diminishing returns in workloads constrained by dependencies, synchronization, or memory stalls. We use a moderately wide configuration with a width of six as the baseline, and evaluate performance at both narrower and wider widths.

Figure~\ref{fig:width_sensitivity} shows the execution time change relative to the baseline. Overall, decreasing width leads to consistent slowdowns across all workloads. For example, \texttt{co} and \texttt{ar} show 37.1\% and 26.4\% longer execution times at width=2, while \texttt{dm} and \texttt{rj} see very minimal slowdowns in execution time of 3.5\% and 3.8\%, respectively. This suggests that some workloads benefit from wide frontends due to their regular numerical kernels or shorter dependency chains, while others are limited by other bottlenecks. Increasing width to 8, on the other hand, yields only marginal improvements, typically less than 2\%. These results indicate that most FEBio workloads do not have enough ILP to saturate wider issue paths and are instead bottlenecked by backend stalls or memory latency. The muted gains from wider pipeline widths reflect the limited ILP observed in the execution stage profile (Figure~\ref{fig:char_execute}). Workloads such as \texttt{rj} and \texttt{dm} exhibit high proportions of branch and memory operations and relatively fewer floating-point and integer instructions, which are typically more parallelizable. In contrast, compute-bound models like \texttt{ar}, \texttt{co}, \texttt{ma} and \texttt{tu}, with a higher share of arithmetic instructions, especially floating-point instructions, benefit more from wider pipelines. This disparity highlights that pipeline width scalability is closely tied to the available mix of independent compute instructions. 
% To this end, a balanced design with 4–6 width stages captures most of the performance benefit with minimal hardware overhead—making it an efficient configuration for domain-specific accelerators targeting biomechanics workloads.

\begin{figure}[!t]
    \centering
    \includegraphics[width=\linewidth]{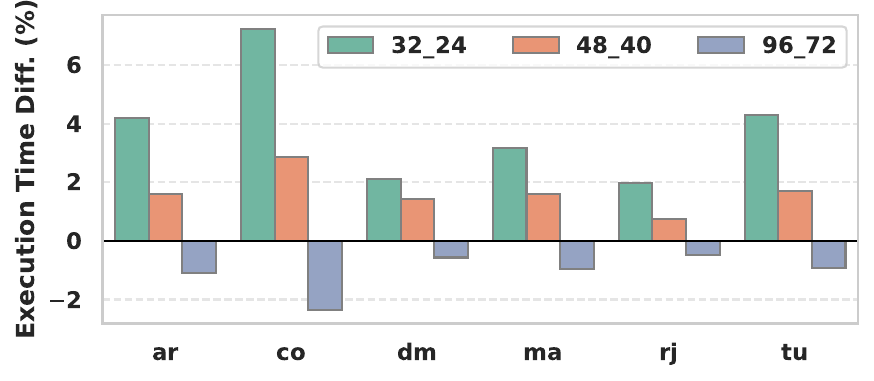}
    \caption{Execution time difference relative to baseline memory queue depths LQ\_SQ = 72\_56.}
    \vspace{-5pt}
    \label{fig:lq_sq_sensitivity}
\end{figure}

\subsubsection{Sensitivity to Memory Queue Depths}

Load and store queues play a critical role in managing memory-level parallelism (MLP) and tolerating latency from cache misses and dependent loads. To evaluate their effect on performance, we vary the number of entries in the load and store queues while keeping all other microarchitectural parameters fixed. This sensitivity study complements earlier findings where backend-bound behavior and memory stalls were observed, especially in irregular or synchronization-heavy models.

Figure~\ref{fig:lq_sq_sensitivity} shows the execution time differences relative to a baseline configuration with 72-entry load queue and 56-entry store queue. Though not as sensitive as pipeline widths, reducing the queue sizes also degrades performance across all workloads. For example, shrinking to \texttt{32\_24} entries results in slowdowns of over 4\% for \texttt{ar}, \texttt{co}, and \texttt{tu}, with \texttt{co} showing the most sensitivity at 7.2\%. These trends suggest that queue pressure is more prominent in workloads with deep dependency chains or high memory concurrency. In contrast, increasing to 96\_72 entries yields only marginal improvements, typically under 2.5\%, indicating that the baseline queue sizes are already large enough to absorb most memory-level concurrency in these workloads. Thus, while undersizing the LQ\_SQ can meaningfully harm performance, oversizing yields little benefit. These results are consistent with the execution and commit stage profiles (Figures~\ref{fig:char_execute},~\ref{fig:char_commit}), where models such as \texttt{co}, \texttt{tu}, and \texttt{ma} show high memory operation intensity. Such characteristics increase pressure on the memory queues and benefit from deeper LQ/SQ buffers. We also experimented with increasing reorder buffer and issue queue sizes, but observed less than 4\% improvement in execution time across workloads. This suggests that instruction windowing is not a dominant limiter, due to moderate ILP and frequent backend stalls that prevent deep instruction reordering from being fully exploited.

\setlength{\textfloatsep}{20pt}

\begin{figure}[!t]
    \centering
    \includegraphics[width=\linewidth]{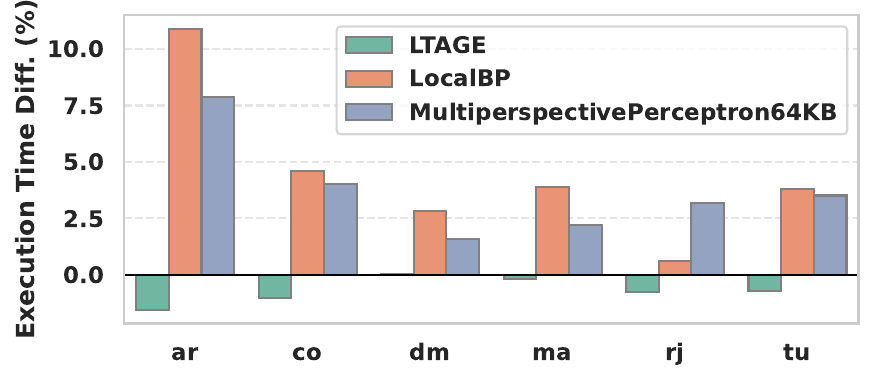}
    \caption{Execution time difference relative to TournamentBP branch predictor.}
    \vspace{-15pt}
    \label{fig:branchpred_percent_diff_plot}
\end{figure}

\subsubsection{Sensitivity to Branch 
Predictors} We evaluate the impact of branch predictor design on execution time by sweeping across four predictor types: LocalBP, TournamentBP (baseline), LTAGE, and MultiperspectivePerceptron. 
% This study is motivated by VTune profiles that highlight moderate frontend stalls and control dependencies in some workloads. 
Our goal is to assess whether advanced predictors yield meaningful gains in scientific computing workloads and whether lightweight alternatives could suffice for efficient domain-specific accelerator (DSA) implementation.

\setlength{\parskip}{5pt} 
   
As shown in Figure~\ref{fig:branchpred_percent_diff_plot}, predictor choice has a measurable but generally modest impact. LTAGE consistently performs best, achieving the lowest execution time in five out of six workloads. Compared to TournamentBP, LTAGE yields $\sim$1.5\% gains in \texttt{ar} but maintains parity or better across all models. In contrast, LocalBP shows the highest sensitivity, performing up to 11\% worse in \texttt{ar}, and 4–5\% worse in \texttt{co} and \texttt{ma}. MultiperspectivePerceptron underperforms LTAGE and TournamentBP in all cases, with overheads up to 7.9\%. These results suggest that while some workloads benefit modestly from deeper or history-based predictors like LTAGE, the overall sensitivity to branch prediction is low relative to backend bottlenecks. This complements the fetch-stage data (Figure~\ref{fig:char_fetch}), which shows that while most models spend the majority of time in regular fetch cycles, control-related stalls (e.g., squashes, trap handling) remain modest, explaining the overall limited sensitivity to branch predictor design. For DSA design, this implies that high-complexity neural predictors (e.g., perceptrons) offer little advantage in this domain. A compact, hardware-efficient predictor such as LTAGE presents a favorable tradeoff, achieving high accuracy without incurring unnecessary area or power costs.

% \begin{tcolorbox}[enhanced,width=3.4in,
%     drop fuzzy shadow southeast,
%     colframe=gray!10,colback=gray!10]
% \textbf{Takeaway VI:} FEBio workloads exhibit mixed sensitivity to microarchitectural features. Performance is most affected by pipeline width, cache capacity, load/store queue depths, and branch predictor capability, while reorder buffers have a marginal impact.
% % These findings suggest that domain-specific accelerators should prioritize moderate pipeline widths (4–6), lean but sufficient cache hierarchies (32\,kB L1, 512\,kB–1\,MB L2), and ample memory-level parallelism to maximize efficiency across diverse biomechanics models.
% \end{tcolorbox}

\begin{tcolorbox}[enhanced,width=3.4in,
    drop fuzzy shadow southeast,
    colframe=gray!10,colback=gray!10]
\textbf{Takeaway II:} 
FEBio workloads show varied instruction profiles, spending most time on integer and floating-point execution, with some models experiencing frontend stalls like instruction cache misses. Performance is sensitive to pipeline width, cache size, load/store queues, and branch prediction. Limited benefits from frequency scaling underscore diminishing returns from deeper pipelines, while larger caches reduce memory stalls. These findings highlight the need for accelerators to support irregular memory access, sparse data, and domain-specific parallelism through hardware-software co-design.
\end{tcolorbox}

\section{Conclusions \& Discussions}
This work introduced Belenos, the first architectural characterization of biomechanical simulation workloads. Built on finite element analysis, these simulations are essential for biomedical research and clinical applications but are often limited by long runtimes and architectural inefficiencies. Our analysis of FEBio models, including both general workloads and a real-world ocular biomechanics case, identified key bottlenecks stemming from irregular memory access, variable compute intensity, and limited architectural support for iterative simulation pipelines. These challenges directly affect the translational potential of biomechanics. To meet runtime constraints, engineers frequently simplify models by reducing geometric complexity, coarsening meshes, or fixing boundary conditions, which lowers computational cost but also compromises fidelity and clinical relevance. The problem becomes even more pronounced in iterative workflows such as parameter fitting or optimization, where repeated simulations are needed but often impractical due to slow performance.

%%from rebuttal
 We are aware of ongoing HBM analyses on AMD/Xilinx datacenter FPGAs, with opportunities to extend bandwidth profiling across additional workload types. Our methodology, based on the architecture-agnostic Top-Down Microarchitecture Analysis (TMA) taxonomy, remains applicable on AMD systems using VTune in software-sampling mode or AMD uProf. On Arm platforms, FEBio support is limited since its default solver (Intel MKL Pardiso) is x86-only.
% ; while it can run on Apple-silicon Macs through Rosetta emulation and an experimental Arm-native port with Panua Pardiso exists, no official release is yet available, meaning most users still profile on x86.
For FEBio developers, we recommend adding a system profiling step that identifies whether models are core- or memory-bound and suggests optimal configurations, as well as expanding portability through a native Arm build and improving solver integration to better exploit optimized math kernels. 
For users, our model-specific bounds help guide system choices, with core-bound models benefiting from CPUs with high single-thread performance and memory-bound models favoring high-bandwidth memory systems.

%Looking forward, FEBio’s performance bottlenecks could be addressed with domain-aware architectural strategies such as near-memory processing (NMP) and reconfigurable logic to accelerate element-wise and matrix-assembly operations.

%%%%

Our findings underscore the importance of domain-aware architectural strategies. Emerging approaches such as near-memory processing can reduce memory-bound stalls by placing compute closer to sparse matrix data in DRAM. Reconfigurable logic, such as FPGAs, can accelerate element-wise and matrix assembly operations with custom pipelines that handle irregular access and control flow. Task-specific accelerators for sparse matrix assembly and iterative solvers, integrated near memory controllers or within cache hierarchies, also represent a promising direction to alleviate backend bottlenecks. Realizing these improvements will require cross-disciplinary co-design, as the irregularity, sparsity, and nonlinear iteration patterns in biomechanics workloads, especially in FEBio, are not efficiently served by general-purpose architectures. Future work should build on this characterization to guide the development of co-designed accelerators and simulation frameworks, ultimately bridging architectural advances with domain-specific demands to enable faster, more scalable, and clinically relevant biomechanical simulations.

\section*{Acknowledgment}

%TODOOOOOO

We thank the members of the Computer Architecture and Systems Lab at UMD, specifically Donghyeon Joo, Ubaid Bakhtiar, Helya Hosseini, and Sanjali Yadav, for helpful discussions on this work.
We would also like to thank the developers and contributors of FEBio, for the resources that made this work possible, and their outstanding contribution to medical sciences. 
Lastly, we thank the anonymous reviewers for their valuable feedback.
This paper is part of a project supported by the U.S. Department of Energy (DoE), Office of Science, ASCR ECRP DE-SC0024079, awarded to Bahar Asgari, which focuses on designing reconfigurable computing systems and will use the results of Belenos for FEA simulations. Johnson Umeike and Amirmahdi Namjoo are supported by Bahar Asgari’s NSF PPoSS program award under Award Number 2316177.

\bibliographystyle{IEEEtranS}
\bibliography{reference}

% Generated by IEEEtranS.bst, version: 1.12 (2007/01/11)
\begin{thebibliography}{10}
\providecommand{\url}[1]{#1}
\csname url@samestyle\endcsname
\providecommand{\newblock}{\relax}
\providecommand{\bibinfo}[2]{#2}
\providecommand{\BIBentrySTDinterwordspacing}{\spaceskip=0pt\relax}
\providecommand{\BIBentryALTinterwordstretchfactor}{4}
\providecommand{\BIBentryALTinterwordspacing}{\spaceskip=\fontdimen2\font plus
\BIBentryALTinterwordstretchfactor\fontdimen3\font minus \fontdimen4\font\relax}
\providecommand{\BIBforeignlanguage}[2]{{%
\expandafter\ifx\csname l@#1\endcsname\relax
\typeout{** WARNING: IEEEtranS.bst: No hyphenation pattern has been}%
\typeout{** loaded for the language `#1'. Using the pattern for}%
\typeout{** the default language instead.}%
\else
\language=\csname l@#1\endcsname
\fi
#2}}
\providecommand{\BIBdecl}{\relax}
\BIBdecl

\bibitem{ahn2016pim}
J.~Ahn, S.~Hong, S.~Yoo, O.~Mutlu, and K.~Choi, ``A scalable processing-in-memory accelerator for parallel graph processing,'' in \emph{Proceedings of the 42nd Annual International Symposium on Computer Architecture (ISCA)}, 2015, pp. 105--117.

\bibitem{arias2011suitable}
J.~Arias-Garc{\'\i}a, R.~P. Jacobi, C.~H. Llanos, and M.~Ayala-Rinc{\'o}n, ``A suitable fpga implementation of floating-point matrix inversion based on gauss-jordan elimination,'' in \emph{2011 vii southern conference on programmable logic (SPL)}.\hskip 1em plus 0.5em minus 0.4em\relax IEEE, 2011, pp. 263--268.

\bibitem{asgari2022fafnir}
B.~Asgari \emph{et~al.}, ``Fafnir: An open-source framework for performance analysis of sparse matrix multiplication on spatial architectures,'' in \emph{Proceedings of the ACM/SIGDA International Symposium on Field-Programmable Gate Arrays (FPGA)}, 2022, pp. 1--10.

\bibitem{asgari2021copernicus}
B.~Asgari, R.~Hadidi, J.~Dierberger, C.~Steinichen, A.~Marfatia, and H.~Kim, ``Copernicus: Characterizing the performance implications of compression formats used in sparse workloads,'' in \emph{2021 IEEE International Symposium on Workload Characterization (IISWC)}.\hskip 1em plus 0.5em minus 0.4em\relax IEEE, 2021, pp. 1--12.

\bibitem{asgari2020alrescha}
B.~Asgari, R.~Hadidi, T.~Krishna, H.~Kim, and S.~Yalamanchili, ``Alrescha: A lightweight reconfigurable sparse-computation accelerator,'' in \emph{2020 IEEE International Symposium on High Performance Computer Architecture (HPCA)}.\hskip 1em plus 0.5em minus 0.4em\relax IEEE, 2020, pp. 249--260.

\bibitem{bakhtiar2024acamar}
U.~Bakhtiar, H.~Hosseini, and B.~Asgari, ``Acamar: A dynamically reconfigurable scientific computing accelerator for robust convergence and minimal resource utilization,'' in \emph{Proceedings of the 57th IEEE/ACM International Symposium on Microarchitecture (MICRO)}, 2024, pp. 1601--1616.

\bibitem{bakhtiar2024pipirima}
U.~Bakhtiar, D.~Joo, and B.~Asgari, ``Pipirima: Predicting patterns in sparsity to accelerate matrix algebra,'' in \emph{Proceedings of the 62nd ACM/IEEE Design Automation Conference (DAC)}, 2025.

\bibitem{balevic2008accelerating}
A.~Balevic, L.~Rockstroh, A.~Tausendfreund, S.~Patzelt, G.~Goch, and S.~Simon, ``Accelerating simulations of light scattering based on finite-difference time-domain method with general purpose gpus,'' in \emph{2008 11th IEEE International Conference on Computational Science and Engineering}.\hskip 1em plus 0.5em minus 0.4em\relax IEEE, 2008, pp. 327--334.

\bibitem{gem5}
\BIBentryALTinterwordspacing
N.~Binkert, B.~Beckmann, G.~Black, S.~K. Reinhardt, A.~Saidi, A.~Basu, J.~Hestness, D.~R. Hower, T.~Krishna, S.~Sardashti, R.~Sen, K.~Sewell, M.~Shoaib, N.~Vaish, M.~D. Hill, and D.~A. Wood, ``The gem5 simulator,'' \emph{SIGARCH Computer Architecture News}, vol.~39, no.~2, pp. 1--7, Aug. 2011. [Online]. Available: \url{https://doi.org/10.1145/2024716.2024718}
\BIBentrySTDinterwordspacing

\bibitem{cao2011three}
L.~Cao, F.~Guilak, and L.~A. Setton, ``Three-dimensional finite element modeling of pericellular matrix and cell mechanics in the nucleus pulposus of the intervertebral disk based on in situ morphology,'' \emph{Biomechanics and modeling in mechanobiology}, vol.~10, pp. 1--10, 2011.

\bibitem{chen20201}
T.~Chen, J.~Botimer, and T.~Chou, ``A 1.87 mm 2 56.9 gops accelerator for solving partial differential equations,'' \emph{IEEE Journal of Solid-State Circuits}, vol.~55, no.~6, pp. 1709--1718, 2020.

\bibitem{chen2019sram}
T.~Chen, J.~Botimer, T.~Chou, and Z.~Zhang, ``An sram-based accelerator for solving partial differential equations,'' in \emph{2019 IEEE Custom Integrated Circuits Conference (CICC)}.\hskip 1em plus 0.5em minus 0.4em\relax IEEE, 2019, pp. 1--4.

\bibitem{chen2024convstencil}
Y.~Chen, K.~Li, Y.~Wang, D.~Bai, L.~Wang, L.~Ma, L.~Yuan, Y.~Zhang, T.~Cao, and M.~Yang, ``Convstencil: Transform stencil computation to matrix multiplication on tensor cores,'' in \emph{Proceedings of the 29th ACM SIGPLAN Annual Symposium on Principles and Practice of Parallel Programming}, 2024, pp. 333--347.

\bibitem{cloutier_pause}
F.~Cloutier, ``Pause — delay execution for a short interval,'' \url{https://www.felixcloutier.com/x86/pause}, 2024, accessed: 2025-06-23.

\bibitem{intel2023vtune}
I.~Corporation, ``{Intel VTune Profiler (Version 2023)},'' \url{https://www.intel.com/content/www/us/en/developer/tools/oneapi/vtune.html}, 2023, available at: \url{https://www.intel.com/content/www/us/en/developer/tools/oneapi/vtune.html}, Accessed: Apr. 4, 2025.

\bibitem{dziekonski2012finite}
A.~Dziekonski, P.~Sypek, A.~Lamecki, and M.~Mrozowski, ``Finite element matrix generation on a gpu,'' \emph{Progress In Electromagnetics Research}, vol. 128, pp. 249--265, 2012.

\bibitem{eggermont2018can}
F.~Eggermont, L.~Derikx, N.~Verdonschot, I.~Van Der~Geest, M.~De~Jong, A.~Snyers, Y.~Van Der~Linden, and E.~Tanck, ``Can patient-specific finite element models better predict fractures in metastatic bone disease than experienced clinicians?: Towards computational modelling in daily clinical practice,'' \emph{Bone \& joint research}, vol.~7, no.~6, pp. 430--439, 2018.

\bibitem{febioGithub}
{FEBio Software}, ``{FEBio} [computer software],'' \url{https://github.com/febiosoftware/FEBio}, 2025, available at: \url{https://github.com/febiosoftware/FEBio}.

\bibitem{feinberg2018enabling}
B.~Feinberg, U.~K.~R. Vengalam, N.~Whitehair, S.~Wang, and E.~Ipek, ``Enabling scientific computing on memristive accelerators,'' in \emph{The International Symposium on Computer Architecture (ISCA)}.\hskip 1em plus 0.5em minus 0.4em\relax IEEE, 2018, pp. 367--382.

\bibitem{feldmannazul}
A.~Feldmann \emph{et~al.}, ``Azul: An accelerator for sparse iterative solvers leveraging distributed on-chip memory,'' in \emph{Proceedings of the 57th IEEE/ACM International Symposium on Microarchitecture (MICRO)}, Nov. 2024.

\bibitem{feldmann2024azul}
A.~Feldmann, C.~Golden, Y.~Yang, J.~S. Emer, and D.~Sanchez, ``Azul: An accelerator for sparse iterative solvers leveraging distributed on-chip memory,'' in \emph{2024 57th IEEE/ACM International Symposium on Microarchitecture (MICRO)}.\hskip 1em plus 0.5em minus 0.4em\relax IEEE, 2024, pp. 643--656.

\bibitem{feng2024efficient}
Y.~Feng, Z.~Sun, C.~Wang, X.~Guo, J.~Mei, Y.~Qi, J.~Liu, J.~Zhang, J.~Wu, X.~Zhan \emph{et~al.}, ``An efficient flash-based computing-in-memory (cim) demonstration of high-precision (32-bit) nonlinear partial differential equation (pde) solver with ultra-high endurance and reliability,'' \emph{IEEE Transactions on Circuits and Systems I: Regular Papers}, 2024.

\bibitem{feola2016finite}
A.~J. Feola, J.~G. Myers, J.~Raykin, L.~Mulugeta, E.~S. Nelson, B.~C. Samuels, and C.~R. Ethier, ``Finite element modeling of factors influencing optic nerve head deformation due to intracranial pressure,'' \emph{Investigative ophthalmology \& visual science}, vol.~57, no.~4, pp. 1901--1911, 2016.

\bibitem{fu2014architecting}
Z.~Fu, T.~J. Lewis, R.~M. Kirby, and R.~T. Whitaker, ``Architecting the finite element method pipeline for the gpu,'' \emph{Journal of computational and applied mathematics}, vol. 257, pp. 195--211, 2014.

\bibitem{fujiki2019transformation}
D.~Fujiki \emph{et~al.}, ``Near-memory data transformation for efficient sparse matrix multi-vector multiplication,'' in \emph{Proceedings of the International Conference for High Performance Computing, Networking, Storage and Analysis (SC)}, 2019, pp. 1--17.

\bibitem{garcia2018step}
E.~Garc{\'\i}a, Y.~Diez, O.~Diaz, X.~Llad{\'o}, R.~Mart{\'\i}, J.~Mart{\'\i}, and A.~Oliver, ``A step-by-step review on patient-specific biomechanical finite element models for breast mri to x-ray mammography registration,'' \emph{Medical physics}, vol.~45, no.~1, pp. e6--e31, 2018.

\bibitem{gerami2024gust}
A.~Gerami and B.~Asgari, ``Gust: Graph edge-coloring utilization for accelerating sparse matrix vector multiplication,'' in \emph{Proceedings of the 29th ACM International Conference on Architectural Support for Programming Languages and Operating Systems, Volume 4}, 2024, pp. 127--141.

\bibitem{giannoula2022sparsep}
C.~Giannoula, I.~Fernandez, J.~G. Luna, N.~Koziris, G.~Goumas, and O.~Mutlu, ``Sparsep: Towards efficient sparse matrix vector multiplication on real processing-in-memory architectures,'' \emph{Proceedings of the ACM on Measurement and Analysis of Computing Systems}, vol.~6, no.~1, pp. 1--49, 2022.

\bibitem{giles2014gpu}
M.~Giles, E.~L{\'a}szl{\'o}, I.~Reguly, J.~Appleyard, and J.~Demouth, ``Gpu implementation of finite difference solvers,'' in \emph{2014 Seventh Workshop on High Performance Computational Finance}.\hskip 1em plus 0.5em minus 0.4em\relax IEEE, 2014, pp. 1--8.

\bibitem{gokhale2015rearrangement}
M.~Gokhale \emph{et~al.}, ``Near memory data structure rearrangement,'' in \emph{Proceedings of the 2015 International Symposium on Memory Systems (MEMSYS)}, 2015, pp. 283--290.

\bibitem{gomez2022benchmarking}
J.~G{\'o}mez-Luna, I.~El~Hajj, I.~Fernandez, C.~Giannoula, G.~F. Oliveira, and O.~Mutlu, ``Benchmarking a new paradigm: Experimental analysis and characterization of a real processing-in-memory system,'' \emph{IEEE Access}, vol.~10, pp. 52\,565--52\,608, 2022.

\bibitem{guo2016energy}
N.~Guo, Y.~Huang, T.~Mai, S.~Patil, C.~Cao, M.~Seok, S.~Sethumadhavan, and Y.~Tsividis, ``Energy-efficient hybrid analog/digital approximate computation in continuous time,'' \emph{IEEE Journal of Solid-State Circuits}, vol.~51, no.~7, pp. 1514--1524, 2016.

\bibitem{han2025flashfftstencil}
H.~Han, K.~Li, W.~Cui, D.~Bai, Y.~Zhang, L.~Yuan, Y.~Chen, Y.~Zhang, T.~Cao, and M.~Yang, ``Flashfftstencil: Bridging fast fourier transforms to memory-efficient stencil computations on tensor core units,'' in \emph{Proceedings of the 30th ACM SIGPLAN Annual Symposium on Principles and Practice of Parallel Programming}, 2025, pp. 355--368.

\bibitem{packer}
{HashiCorp}, ``{Packer},'' \url{https://www.packer.io/}, 2024, version 1.10.0. Accessed June 2025.

\bibitem{hegde2019extensor}
K.~Hegde \emph{et~al.}, ``Extensor: An accelerator for sparse tensor algebra,'' in \emph{Proceedings of the 52nd Annual IEEE/ACM International Symposium on Microarchitecture (MICRO)}, 2019, pp. 319--333.

\bibitem{holanda2011fpga}
B.~Holanda, R.~Pimentel, J.~Barbosa, R.~Camarotti, A.~Silva-Filho, L.~Joao, V.~Souza, J.~Ferraz, and M.~Lima, ``An fpga-based accelerator to speed-up matrix multiplication of floating point operations,'' in \emph{2011 IEEE International Symposium on Parallel and Distributed Processing Workshops and Phd Forum}.\hskip 1em plus 0.5em minus 0.4em\relax IEEE, 2011, pp. 306--309.

\bibitem{hosseini2025segin}
H.~Hosseini, U.~Bakhtiar, D.~Joo, and B.~Asgari, ``Segin: Synergistically enabling fine-grained multi-tenant and resource optimized spmv,'' \emph{IEEE Computer Architecture Letters}, 2025.

\bibitem{huang2017hybrid}
Y.~Huang, N.~Guo, M.~Seok, Y.~Tsividis, K.~Mandli, and S.~Sethumadhavan, ``Hybrid analog-digital solution of nonlinear partial differential equations,'' in \emph{MICRO}.\hskip 1em plus 0.5em minus 0.4em\relax IEEE, 2017, pp. 665--678.

\bibitem{huthwaite2014accelerated}
P.~Huthwaite, ``Accelerated finite element elastodynamic simulations using the gpu,'' \emph{Journal of Computational Physics}, vol. 257, pp. 687--707, 2014.

\bibitem{intel_pardiso}
{Intel Corporation}, \emph{{Intel\textsuperscript{\textregistered} oneAPI Math Kernel Library (oneMKL) PARDISO Solver}}, \url{https://www.intel.com/content/www/us/en/docs/onemkl/developer-reference-c/2023-0/onemkl-pardiso-parallel-direct-sparse-solver-iface.html}, 2023, available at: \url{https://www.intel.com/content/www/us/en/docs/onemkl/developer-reference-c/2023-0/onemkl-pardiso-parallel-direct-sparse-solver-iface.html}.

\bibitem{intel_mkl}
\BIBentryALTinterwordspacing
{Intel Corporation, oneMKL Team}, \emph{Intel\textsuperscript{\textregistered} Math Kernel Library}, 2024, version 2024. [Online]. Available: \url{https://www.intel.com/content/www/us/en/developer/tools/oneapi/onemkl.html}
\BIBentrySTDinterwordspacing

\bibitem{irturk2008fpga}
A.~Irturk, B.~Benson, S.~Mirzaei, and R.~Kastner, ``An fpga design space exploration tool for matrix inversion architectures,'' in \emph{2008 Symposium on Application Specific Processors}.\hskip 1em plus 0.5em minus 0.4em\relax IEEE, 2008, pp. 42--47.

\bibitem{jacquelin2022scalable}
M.~Jacquelin, M.~Araya-Polo, and J.~Meng, ``Scalable distributed high-order stencil computations,'' in \emph{2022 SC22: International Conference for High Performance Computing, Networking, Storage and Analysis (SC)}.\hskip 1em plus 0.5em minus 0.4em\relax IEEE Computer Society, 2022, pp. 411--423.

\bibitem{karkooti2005fpga}
M.~Karkooti, J.~R. Cavallaro, and C.~Dick, ``Fpga implementation of matrix inversion using qrd-rls algorithm,'' in \emph{Asilomar Conference on Signals, Systems, and Computers}, 2005.

\bibitem{ke2020recnmp}
L.~Ke \emph{et~al.}, ``Recnmp: Accelerating personalized recommendation with near-memory processing,'' in \emph{Proceedings of the 47th Annual International Symposium on Computer Architecture (ISCA)}, 2020, pp. 313--326.

\bibitem{khoram2017challenges}
S.~Khoram, H.~Zhan, A.~Majumdar, J.~Li, Z.~Qian, and L.~Song, ``Challenges and opportunities: From near-memory computing to in-memory computing,'' in \emph{Proceedings of the 2017 ACM on International Symposium on Physical Design (ISPD '17)}, 2017, pp. 43--46.

\bibitem{kopysov2016scalability}
S.~Kopysov, A.~Novikov, N.~Nedozhogin, and V.~Rychkov, ``Scalability of parallel finite element algorithms on multi-core platforms,'' in \emph{IOP Conference Series: Materials Science and Engineering}, vol. 158, no.~1.\hskip 1em plus 0.5em minus 0.4em\relax IOP Publishing, 2016, p. 012055.

\bibitem{kwon2019tensordimm}
Y.~Kwon \emph{et~al.}, ``Tensordimm: A practical near-memory processing architecture for embeddings and tensor operations in deep learning,'' in \emph{Proceedings of the 52nd Annual IEEE/ACM International Symposium on Microarchitecture (MICRO)}, 2019, pp. 740--753.

\bibitem{lee2021vivo}
C.~Lee, G.~Li, W.~D. Stamer, and C.~R. Ethier, ``In vivo estimation of murine iris stiffness using finite element modeling,'' \emph{Experimental eye research}, vol. 202, p. 108374, 2021.

\bibitem{li2024aging}
G.~Li, J.~van Batenburg-Sherwood, B.~N. Safa, N.~S. Fraticelli~Guzm{\'a}n, A.~Wilson, M.~R. Bahrani~Fard, K.~Choy, M.~L. De~Ieso, J.~S. Cui, A.~J. Feola \emph{et~al.}, ``Aging and intraocular pressure homeostasis in mice,'' \emph{Aging Cell}, vol.~23, no.~7, p. e14160, 2024.

\bibitem{li2023fdmax}
J.~Li, Y.~Zhang, H.~Zheng, and K.~Wang, ``Fdmax: An elastic accelerator architecture for solving partial differential equations,'' in \emph{Proceedings of the 50th Annual International Symposium on Computer Architecture}, 2023, pp. 1--12.

\bibitem{lin2013design}
C.-Y. Lin and I.~Verbauwhede, ``Design space exploration for sparse matrix-matrix multiplication on fpgas,'' \emph{International Journal of Circuit Theory and Applications}, vol.~41, no.~2, pp. 199--211, 2013.

\bibitem{lloyd2017lookup}
S.~Lloyd and M.~Gokhale, ``Near memory key/value lookup acceleration,'' in \emph{Proceedings of the International Symposium on Memory Systems (MEMSYS)}, 2017, pp. 26--33.

\bibitem{ltaief2023scaling}
H.~Ltaief, Y.~Hong, L.~Wilson, M.~Jacquelin, M.~Ravasi, and D.~E. Keyes, ``Scaling the “memory wall” for multi-dimensional seismic processing with algebraic compression on cerebras cs-2 systems,'' in \emph{Proceedings of the International Conference for High Performance Computing, Networking, Storage and Analysis}, 2023, pp. 1--12.

\bibitem{maas2012febio}
S.~A. Maas, B.~J. Ellis, G.~A. Ateshian, and J.~A. Weiss, ``Febio: Finite elements for biomechanics,'' \emph{Journal of Biomechanics}, vol.~45, no.~6, pp. 1102--1105, 2012.

\bibitem{maas2009comparison}
S.~A. Maas, B.~J. Ellis, and J.~A. Weiss, ``A comparison of febio, abaqus, and nike3d: Results for a suite of verification problems,'' The University of Utah, Salt Lake City, UT, USA, Tech. Rep., 2009.

\bibitem{markall2010towards}
G.~R. Markall, D.~A. Ham, and P.~H. Kelly, ``Towards generating optimised finite element solvers for gpus from high-level specifications,'' \emph{Procedia Computer Science}, vol.~1, no.~1, pp. 1815--1823, 2010.

\bibitem{martinez2015efficient}
J.~Mart{\'\i}nez-Frutos and D.~Herrero-P{\'e}rez, ``Efficient matrix-free gpu implementation of fixed grid finite element analysis,'' \emph{Finite Elements in Analysis and Design}, vol. 104, pp. 61--71, 2015.

\bibitem{merema2021patient}
B.~B.~J. Merema, J.~Kraeima, H.~H. Glas, F.~K. Spijkervet, and M.~J. Witjes, ``Patient-specific finite element models of the human mandible: Lack of consensus on current set-ups,'' \emph{Oral diseases}, vol.~27, no.~1, pp. 42--51, 2021.

\bibitem{michea2010accelerating}
D.~Mich{\'e}a and D.~Komatitsch, ``Accelerating a three-dimensional finite-difference wave propagation code using gpu graphics cards,'' \emph{Geophysical Journal International}, vol. 182, no.~1, pp. 389--402, 2010.

\bibitem{micikevicius20093d}
P.~Micikevicius, ``3d finite difference computation on gpus using cuda,'' in \emph{Proceedings of 2nd workshop on general purpose processing on graphics processing units}, 2009, pp. 79--84.

\bibitem{morgan2016finite}
A.~E. Morgan, J.~L. Pantoja, J.~Weinsaft, E.~Grossi, J.~M. Guccione, L.~Ge, and M.~Ratcliffe, ``Finite element modeling of mitral valve repair,'' \emph{Journal of biomechanical engineering}, vol. 138, no.~2, p. 021009, 2016.

\bibitem{mu2024scalable}
J.~Mu, C.~Yu, T.~T.-H. Kim, and B.~Kim, ``A scalable and reconfigurable bit-serial compute-near-memory hardware accelerator for solving 2-d/3-d partial differential equations,'' \emph{IEEE Journal of Solid-State Circuits}, 2024.

\bibitem{nurvitadhi2015sparse}
E.~Nurvitadhi, G.~Venkatesh, D.~Marr, R.~Huang, J.~Ong, J.~MacKean, V.~Talwar, and D.~D. Goehringer, ``A sparse matrix vector multiply accelerator for support vector machine,'' in \emph{Proceedings of the 2015 International Conference on Compilers, Architecture and Synthesis for Embedded Systems (CASES)}.\hskip 1em plus 0.5em minus 0.4em\relax IEEE Press, 2015, pp. 1--10.

\bibitem{okimura2013parallelization}
T.~Okimura, T.~Sasayama, N.~Takahashi, and S.~Ikuno, ``Parallelization of finite element analysis of nonlinear magnetic fields using gpu,'' \emph{IEEE transactions on magnetics}, vol.~49, no.~5, pp. 1557--1560, 2013.

\bibitem{openmp}
{OpenMP Architecture Review Board}, ``Openmp application programming interface,'' \url{https://www.openmp.org}, 2025, available at: \url{https://www.openmp.org}.

\bibitem{ozer2025superba}
N.~Ozer, G.~Kollmer, R.~Hadidi, and B.~Asgari, ``La superba: Leveraging a self-comparison method to understand the performance benefits of sparse acceleration optimizations,'' in \emph{2025 IEEE International Symposium on Performance Analysis of Systems and Software (ISPASS)}.\hskip 1em plus 0.5em minus 0.4em\relax IEEE, 2025, pp. 1--12.

\bibitem{pal2018outerspace}
S.~Pal, A.~Bhowmick, and A.~Chattopadhyay, ``Outerspace: An outer product based sparse matrix multiplication accelerator,'' in \emph{Proceedings of the 2018 IEEE International Symposium on High Performance Computer Architecture (HPCA)}.\hskip 1em plus 0.5em minus 0.4em\relax IEEE, 2018, pp. 724--735.

\bibitem{peverelli2022characterizing}
F.~Peverelli, F.~Zaruba, and L.~Benini, ``Characterizing molecular dynamics simulation on commodity platforms,'' in \emph{2022 IEEE International Symposium on Workload Characterization (IISWC)}, 2022, pp. 65--78.

\bibitem{ramchandani2023spica}
D.~Ramchandani, B.~Asgari, and H.~Kim, ``Spica: Exploring fpga optimizations to enable an efficient spmv implementation for computations at edge,'' in \emph{2023 IEEE International Conference on Edge Computing and Communications (EDGE)}.\hskip 1em plus 0.5em minus 0.4em\relax IEEE, 2023, pp. 36--42.

\bibitem{rocki2020fast}
K.~Rocki, D.~Van~Essendelft, I.~Sharapov, R.~Schreiber, M.~Morrison, V.~Kibardin, A.~Portnoy, J.~F. Dietiker, M.~Syamlal, and M.~James, ``Fast stencil-code computation on a wafer-scale processor,'' in \emph{SC20: International Conference for High Performance Computing, Networking, Storage and Analysis}.\hskip 1em plus 0.5em minus 0.4em\relax IEEE, 2020, pp. 1--14.

\bibitem{rodrigues2019scatter}
A.~Rodrigues \emph{et~al.}, ``Towards a scatter-gather architecture: Hardware and software issues,'' in \emph{Proceedings of the International Symposium on Memory Systems (MEMSYS)}, 2019, pp. 261--271.

\bibitem{rodriguez2019blas}
E.~Rodriguez-Gutiez, A.~Moreton-Fernandez, A.~Gonzalez-Escribano \emph{et~al.}, ``Toward a blas library truly portable across different accelerator types,'' \emph{Journal of Supercomputing}, vol.~75, pp. 7101--7124, 2019.

\bibitem{ross2024bayesian}
C.~J. Ross, D.~W. Laurence, A.~Aggarwal, M.-C. Hsu, A.~Mir, H.~M. Burkhart, and C.-H. Lee, ``Bayesian optimization-based inverse finite element analysis for atrioventricular heart valves,'' \emph{Annals of Biomedical Engineering}, vol.~52, no.~3, pp. 611--626, 2024.

\bibitem{safa2023effects}
B.~N. Safa, A.~J. Razeghinejad, J.~A. Downs, H.~Reynaud, T.~A. Swindle-Reilly, and M.~R. Eskandari, ``The effects of negative periocular pressure on biomechanics of the optic nerve head and cornea: A computational modeling study,'' \emph{Translational Vision Science \& Technology}, vol.~12, no.~2, p.~5, 2023.

\bibitem{safa2020evaluation}
B.~N. Safa, E.~T. Bloom, A.~H. Lee, M.~H. Santare, and D.~M. Elliott, ``Evaluation of transverse poroelastic mechanics of tendon using osmotic loading and biphasic mixture finite element modeling,'' \emph{Journal of biomechanics}, vol. 109, p. 109892, 2020.

\bibitem{safa2019helical}
B.~N. Safa, J.~M. Peloquin, J.~R. Natriello, J.~L. Caplan, and D.~M. Elliott, ``Helical fibrillar microstructure of tendon using serial block-face scanning electron microscopy and a mechanical model for interfibrillar load transfer,'' \emph{Journal of the Royal Society Interface}, vol.~16, no. 160, p. 20190547, 2019.

\bibitem{safa2021}
E.~C. E.~D. Safa~BN, Santare~MH, ``Identifiability of tissue material parameters from uniaxial tests using multi-start optimization.'' \emph{Acta Biomaterialia}, vol. 123, pp. 197--207, 2021.

\bibitem{singh2020nero}
G.~Singh, D.~Diamantopoulos, C.~Hagleitner, J.~G{\'o}mez-Luna, S.~Stuijk, O.~Mutlu, and H.~Corporaal, ``Nero: A near high-bandwidth memory stencil accelerator for weather prediction modeling,'' in \emph{2020 30th International Conference on Field-Programmable Logic and Applications (FPL)}.\hskip 1em plus 0.5em minus 0.4em\relax IEEE, 2020, pp. 9--17.

\bibitem{febioStudioUserManual}
F.~Software, \emph{{FEBioStudio 1.7 User Manual}}, \url{https://help.febio.org/docs/FEBioStudio-1-7/FSM17.html}, 2021, available at: \url{https://help.febio.org/docs/FEBioStudio-1-7/FSM17.html}.

\bibitem{sun2007implementation}
Q.~Sun, M.~Zhao, and X.~Zhang, ``Implementation matrix inversion of lu algorithm with systolic array,'' \emph{Microelectronics and Computers}, vol.~24, pp. 138--141, 2007.

\bibitem{tanabe2011gather}
N.~Tanabe, K.~Ishizaki, and H.~Amano, ``A memory accelerator with gather functions for bandwidth-bound irregular applications,'' in \emph{Proceedings of the 1st Workshop on Irregular Applications: Architectures and Algorithms (IA3)}, 2011, pp. 35--42.

\bibitem{vadlamudi2024electra}
C.~K. Vadlamudi and B.~Asgari, ``Electra: Eliminating the ineffectual computations on bitmap compressed matrices,'' \emph{IEEE Computer Architecture Letters}, 2024.

\bibitem{FEBio_Theory_Manual}
J.~A. Weiss \emph{et~al.}, \emph{{FEBio Theory Manual Version 3.6}}, \url{https://febio.org/theory-manual}, 2019, available at: \url{https://febio.org/theory-manual}, Accessed: Apr. 3, 2025.

\bibitem{xie2024fast}
C.~Xie, Z.~Shao, N.~Zhao, X.~Hu, Y.~Du, and L.~Du, ``A fast-convergence near-memory-computing accelerator for solving partial differential equations,'' \emph{IEEE Transactions on Very Large Scale Integration (VLSI) Systems}, 2024.

\bibitem{xu2018improved}
Y.~Xu, D.~Li, Y.~Xi, J.~Lan, and T.~Jiang, ``An improved predictive controller on the fpga by hardware matrix inversion,'' \emph{IEEE Transactions on Industrial Electronics}, vol.~65, no.~9, pp. 7395--7405, 2018.

\bibitem{yadav2025dynaflow}
S.~Yadav and B.~Asgari, ``Dynaflow: An ml framework for dynamic dataflow selection in spgemm accelerators,'' \emph{IEEE Computer Architecture Letters}, 2025.

\bibitem{zhang2024lorastencil}
Y.~Zhang, K.~Li, L.~Yuan, J.~Cheng, Y.~Zhang, T.~Cao, and M.~Yang, ``Lorastencil: Low-rank adaptation of stencil computation on tensor cores,'' in \emph{SC24: International Conference for High Performance Computing, Networking, Storage and Analysis}.\hskip 1em plus 0.5em minus 0.4em\relax IEEE, 2024, pp. 1--17.

\end{thebibliography}

% \include{ae_vtune}
% \documentclass{sigplanconf}
 %\usepackage{hyperref}

%\begin{document}

\fontsize{9pt}{11pt}\selectfont

\appendix
\section{Artifact Appendix}

%%%%%%%%%%%%%%%%%%%%%%%%%%%%%%%%%%%%%%%%%%%%%%%%%%%%%%%%%%%%%%%%%%%%%
\subsection{Abstract}

% This artifact provides the necessary components to reproduce the performance profiling and analysis of biomechanical simulations using FEBio, as presented in the paper. It includes datasets from the FEBio test suite and a high-resolution ocular biomechanics case study, along with instructions for running FEBio simulations, analyzing them with Intel VTune Profiler (part 1), and compiling the gem5-based microarchitectural evaluation (part 2). We release (i) scripts to build a bootable disk image and Linux kernel, (ii) automated experiment launchers that sweep frequency, cache sizes, branch predictors, and key pipeline parameters, and (iii) precomputed results plus plotting scripts that regenerate the paper figures (Figs.~7--12). Evaluators can either re-run simulations (A) or reproduce plots from packaged results (B).

This artifact contains all components needed to reproduce the performance profiling and analysis of biomechanical simulations with FEBio. It includes datasets from the FEBio test suite and an ocular biomechanics case study, plus instructions for running FEBio with Intel VTune Profiler (part 1) and compiling the gem5-based microarchitectural evaluation (part 2). We provide (i) scripts to build a bootable disk image and Linux kernel, (ii) automated experiment launchers for sweeping frequency, cache, branch predictor, and pipeline parameters, and (iii) precomputed results with plotting scripts to regenerate Figs.~7--12. Evaluators can either re-run simulations (A) or reproduce plots from packaged results (B).

%%%%%%%%%%%%%%%%%%%%%%%%%%%%%%%%%%%%%%%%%%%%%%%%%%%%%%%%%%%%%%%%%%%%%

\subsection{Artifact check-list -- Part 1}

{\small
\begin{itemize}
  \setlength\itemindent{-1.5em}
  %\item \textbf{Algorithm:} % leave generic
  
  \item \textbf{Program:} \texttt{FEBio} (\texttt{febio4}); Intel VTune Profiler 2024.1  % TODO: add exact FEBio version/commit
  
  \item \textbf{Compilation:} \texttt{FEBio}: CMake + GCC/Clang; MKL/PARDISO % TODO: pin compiler versions 
  
  \item \textbf{Binary:} FEBio executable, VTune CLI
  
  \item \textbf{Data Set:} Uploaded to Zenodo.

  \item \textbf{Run-time environment:} Ubuntu 20.04 LTS; Intel VTune 2024.1; Intel oneMKL; Python 3.10+.
  \item \textbf{Hardware:} Baseline: Intel Core i9-14900K, 64 GB DDR5, single-socket desktop; any x86\_64 with VTune supported.% Preferably Intel CPU.
  
  \item \textbf{Run-time state:} Using Performance Cores; CPU affinity pinned; \texttt{OMP\_NUM\_THREADS=8} % TODO: confirm governor and pinning policy

  \item \textbf{Execution:}
  % \vspace*{-.5em}
  \hspace{1cm}
  \begin{verbatim}
 sudo /opt/intel/oneapi/vtune/latest/bin64/vtune
-collect uarch-exploration -collect memory-access 
-result-dir .result-file-name -- ./bin/febio4
    \end{verbatim}
     \vspace*{-1em}
    \begin{verbatim}
run file-name.feb
    \end{verbatim}
% \vspace*{-1em}
% \vspace{-.5em}
  \item \textbf{Metrics:} VTune Top-Down %(Retiring/FE/BadSpec/BE; FE latency/bandwidth; BE core/memory)
  , CPI, MPKI, hotspot tables 
  
  \item \textbf{Output:} VTune result directories and CSV reports are included in the Zenodo records.
  
  \item \textbf{Experiments:}  11-model set + optional eye model % fixed
  
  \item \textbf{How much disk space required?:} \textasciitilde 2GB

  \item \textbf{How much time to prepare workflow?:} \textasciitilde 30 minutes
  
  \item \textbf{How much time to complete experiments?:} 
   \textasciitilde 1–2 hours (small subset), 18+ hours (full paper results)
   
  \item \textbf{Publicly available?:} 
  Code, Datasets, Scripts: Included in Zenodo files. 
  
  \item \textbf{Code licenses (if public)?:} 
  Automation scripts (Bash/Python): MIT License ;  
  FEBio (solver software): permissive Apache License 2.0.

  \item \textbf{Data licenses (if public)?:} 
  \textbf{FEBio model input files (.feb):} No specific licensing restrictions from FEBio; we release these under CC BY 4.0.
  
  \item \textbf{Workflow framework used?:} Plain shell + Python;  
  
  \item \textbf{Archived (provide DOI)?:} \url{https://doi.org/10.5281/zenodo.16924703}
\end{itemize}
}

%%%%%%%%%%%%%%%%%%%%%%%%%%%
% \vspace{-.5em}
\subsection{Artifact check-list -- Part 2}
{\small
\begin{itemize}
 \setlength\itemindent{-2em}
  \item {\bf Program:} gem5 (X86 FS); FEBio solver in the guest OS
  \item {\bf Compilation:} gcc/clang for host; disk image build in README
  \item {\bf Binary:} gem5.opt, guest \texttt{vmlinux}, \texttt{febio\_software.img}
  \item {\bf Data set:} FEBio models (\texttt{ar, co, dm, ma, rj, tu}); eye model for checkpoint
  \item {\bf Run-time environment:} Linux (Ubuntu LTS is recommended)
  \item {\bf Hardware:} If re-running: $\geq$ 8 CPU cores and 64\,GB RAM
  \item {\bf Execution:} Bash scripts (non-interactive); long-running gem5 when re-running
  \item {\bf Metrics:} Execution time, IPC, MPKI, and stage breakdowns
  \item {\bf Output:} \texttt{stats.txt}, \texttt{simout}, PDFs of plots (Figs.~7--12)
  \item {\bf Experiments:} Frequency, L1/L2 capacity, branch predictor, width \& queue sizes
  \item {\bf Disk space:} $\sim$30–50 GB (re-run); $\sim$1–5 GB (packaged results only)
  \item {\bf Prep time:} $\sim$1--2 h (install deps, build image/kernel)
  \item {\bf Experiment time:} Full gem5 runs can take many hours to days (depending on workload and configuration); reproducing plots from packaged results takes only a few minutes.
  \item {\bf Publicly available?:} Yes (code and results on GitHub/Zenodo) 
  \item {\bf Code licenses:} MIT/BSD-style (project) + gem5 license (upstream)
  \item {\bf Data licenses:} As per FEBio models (redistributed where allowed)
  \item {\bf Workflow automation:} Bash scripts
  \item {\bf Archived (DOI)?:} Zenodo DOI \url{https://doi.org/10.5281/zenodo.16924703}
\end{itemize}
}

%%%%%%%%%%%%%%%%%%%%%%%%%%%%%%%%%%%%%%%%%%%%%%%%%%%%%%%%%%%%%%%%%%%%%
\vspace{-.75em}\subsection{Description}

\subsubsection{How to access -- Part 1}
\begin{itemize}
 \setlength\itemindent{-2em}
  \item \textbf{Artifact bundle %(Part A)
  }: Zenodo (Belenos-Vtune.tar.gz) %containing:
  % \texttt{models/} — FEBio inputs: \texttt{bp07..bp09}, \texttt{ma26..ma31}, \texttt{fl33, fl34}; \texttt{eye/} .
  
  %  ; \texttt{bin/} — optional prebuilt \texttt{febio} (Linux x86\_64) with MKL/PARDISO. %\textbf{TODO: include or not}
    
  \item \textbf{System requirements}: x86\_64 Linux with Intel VTune %2024.1 %; sudo not required. % VTune CLI is sufficient
\end{itemize}
%\vspace{-.75em}
\subsubsection{Hardware dependencies -- Part 1}
Baseline used in paper: Intel Core i9-14900K (24 threads / 32 logical), 64GB DDR5 @6000MHz, Ubuntu 20.04 LTS (Linux 5.15). Other x86\_64 systems are acceptable, but figures may vary. Pin threads and fix CPU freq for reproducibility. % (paper baseline)
% TODO: If you will also test on a laptop/server, add a short table of machines.
%\vspace{-.75em}
\subsubsection{Software dependencies -- Part 1}
GCC/Clang, CMake, Intel oneMKL (PARDISO), OpenMP runtime, Intel VTune 2024.1 (Microarchitecture Exploration), Python 3.10+ with \texttt{pandas}, \texttt{matplotlib}. % TODO: versions
%\vspace{-.75em}
\subsubsection{Data sets -- Part 1}
We use three FEBio test-suite groups with identical mesh sizes but differing configurations: Group 1 (bp07–bp09), Group 2 (ma26–ma31), and Group 3 (fl33, fl34)
 % fixed by paper
% TODO: include brief one-line description per case and model sizes. -- check :)

    % \textbf{Group 1 (bp07–bp09):} Biphasic contact mechanics models of cartilage/tissue interaction, medium-sized meshes.
    
    % \textbf{Group 2 (ma26–ma31):} Arterial wall mechanics models under physiological load, smaller to medium meshes.
    
    % \textbf{Group 3 (fl33, fl34):} Fluid dynamics models of incompressible flow, largest meshes in the benchmark set.

    % The \emph{eye} case is a large, sparse, data-dependent model (long runtime).

%%%%%%%%%%%%%%%%%%%%
%\vspace{-.6em}
\subsubsection{How to access -- Part 2}

Source, scripts, and packaged results in Zenodo (Belenos-gem5.tar.gz). The repository contains:
\begin{itemize}
 \setlength\itemindent{-2em}
  \item \texttt{run-scripts/} (experiment launchers)
  \item \texttt{guest-scripts/} (workload runners)
  \item \texttt{resources/} (\texttt{vmlinux}, and \texttt{febio\_software.img})
  \item \texttt{outdir/main-results/exps/} (packaged results) as well as \texttt{plot-scripts/}
\end{itemize}
%\vspace{-1em}
\subsubsection{Hardware dependencies -- Part 2}
Linux host with $\geq$8 CPU cores and 64 GB RAM for re-running simulations; standard x86\_64.
%\vspace{-.75em}
\subsubsection{Software dependencies  -- Part 2}
Ubuntu LTS (recommended), Python~3 with matplotlib/pandas/numpy, gcc/clang, build tools, and gem5 build prerequisites. Kernel/image build dependencies are listed in \texttt{README.md}.
%\vspace{-.8em}
\subsubsection{Data sets  -- Part 2}
FEBio workloads (\texttt{ar, co, dm, ma, rj, tu}); the eye model script used for checkpointing.
%\vspace{-.8em}
\subsubsection{Models  -- Part 2}
gem5 X86 FS with O3 CPU; sweeps over frequency, L1/L2 sizes, branch predictors, width, and load/store queue sizes.

%%%%%%%%%%%%%%%%%%%%%%%%%%%%%%%%%%%%%%%%%%%%%%%%%%%%%%%%%%%%%%%%%%%%%
\vspace{-1em}
\subsection{Installation — Part 1
}
\begin{enumerate}
 \setlength\itemindent{-2em}
  \item \textbf{Install FEBio} (prebuilt recommended):
    % \begin{verbatim}
    % # TODO: pin compiler versions
    % git clone https://github.com/febiosoftware/FEBio
    % \end{verbatim}
    Using \url{https://febio.org/downloads/} %create an account and download the Linux version of Febio4. Follow the given setup instructions. 

  \item \textbf{Install Intel VTune 2024.1} %following the instructions found here 
  Using \url{https://www.intel.com/content/www/us/en/developer/tools/oneapi/vtune-profiler-download.html?operatingsystem=linux&linux-install-type=offline}.
  \item \textbf{Datasets}: Place \texttt{.feb} files in FEBio's \textit{/bin} directory
\end{enumerate}
%%%%%%%%%%%%%%%%%%%%%%%%%%%%%%
\vspace{-1em}\subsection{Installation -- Part 2}
% \begin{enumerate}
%  \setlength\itemindent{-2em}
  % \item 
  Download the Belenos repository. Then, build gem5 as illustrated in the workflow below.
  % \item (Option A re-run) Build the guest disk image according to \texttt{README.md}.
  % % \item Verify Python~3 environment for plotting: \texttt{pip install -r requirements.txt} (if provided).
 % \end{enumerate}

%%%%%%%%%%%%%%%%%%%%%%%%%%%%%%%%%%%%%%%%%%%%%%%%%%%%%%%%%%%%%%%%%%%%%
\vspace{-1em}
\subsection{Experiment workflow — Part 1
}
\paragraph{Environment.} Set fixed frequency (optional) and threads:
\begin{verbatim}
export OMP_NUM_THREADS=8 # plateau beyond 8 in our tests
\end{verbatim}
%# TODO: optionally set CPU governor to performance; use taskset to pin

% \paragraph{Functional test (one model).}
% \begin{verbatim}
% # Run FEBio without VTune to verify:
% ./bin/febio4 -i models/bp07.feb -o out/bp07.log
% \end{verbatim}

\paragraph{Collect VTune (Microarchitecture Exploration).}
\begin{verbatim}
# Some VTune versions name this analysis 
'uarch-exploration'.
# Use 'vtune -help collect' if needed.
vtune -collect uarch-exploration -collect memory-access
-result-dir .result-file-name -- ./bin/febio4
run file-name.feb
\end{verbatim}

% \paragraph{Batch run (all-model set).}
% \vspace{-.75em}
% \begin{verbatim}
% ./scripts/run_vtune.sh models/list_vtune_all.txt
% # TODO: provide this list file with one path per line:
% # models/bp07.feb ... models/bp09.feb
% # models/ma26.feb ... models/ma31.feb
% # models/fl33.feb models/fl34.feb
% \end{verbatim}

% \paragraph{Export CSV + summaries.}
% \begin{verbatim}
% # Top-Down summary per run:
% vtune -report summary -r vtune_out/bp07 -format csv \
%       -report-output reports/bp07_summary.csv
% \end{verbatim}
%# (Optional) Top-down breakdown report:
% vtune -report topdown -r vtune_out/bp07 -format csv \
%       -report-output reports/bp07_topdown.csv
% /plot_td.py vtune_11.csv figs/

%%%%%%%%%%%%%%%

\subsection{Experiment workflow -- Part 2}
\paragraph{(A) Re-run simulations (optional).}
\begin{enumerate}
 \setlength\itemindent{-2em}
  \item{\bf Install gem5 dependencies:}
\vspace{-0.5em}
\begin{verbatim}
cd ~/Belenos/gem5-exp/
sudo apt update && sudo apt install -y \
build-essential gcc g++ clang python3 \
python3-pip git zlib1g-dev \
libprotobuf-dev protobuf-compiler
\end{verbatim}
\vspace{-0.5em}
\item{\bf Clone and build gem5:} (skip clone if using packaged gem5)  
% \vspace{-0.5em}
\begin{verbatim}
git clone https://github.com/gem5/gem5.git
cd gem5
scons build/X86/gem5.opt -j$(nproc)
\end{verbatim}
% \vspace{-0.5em}
  \item {\bf Build artifacts:} disk image (see \texttt{disk-image/ README.md}). \texttt{vmlinux} is already provided in \texttt{resources/} directory.
  \item {\bf Take checkpoint:}
% \vspace{-0.5em}
\begin{verbatim}
./run-scripts/febio_run_exps.sh \
--num-cpus 2 --script eye_model.sh \
--freq 3GHz --mem-size 4GB \
--take-checkpoint 2>&1  
\end{verbatim}
% \vspace{-0.5em}
  \item {\bf Launch Experiments:}
% \vspace{-0.5em}
\begin{verbatim}
./run-scripts/run_other_configs.sh
./run-scripts/run_micro_configs.sh
\end{verbatim}
% \vspace{-0.5em}
  \item {\bf Plot results:} follow (B) below.
\end{enumerate}

\paragraph{(B) Reproduce plots from packaged results (recommended)}
\begin{enumerate}
\setlength\itemindent{-2em}
  \item \textbf{Change directory to plot scripts:}
  % \vspace{-0.5em}
  \begin{verbatim}
cd ~/Belenos/gem5-exp/outdir/main-results\
/exps/plot-scripts
  \end{verbatim}
  % \vspace{-1.5em}
  \item \textbf{Figs.~7 (pipeline profile: fetch/exec/commit).}
 
  \begin{verbatim}
cd fetch-stage
python3 extract_fetch_data.py && cd ..
cd exec-stage
python3 extract_exec_data.py && cd ..
cd commit-stage
python3 extract_commit_data.py && cd ..
cd pipeline-profile 
python3 plot_pipeline_profile.py
  \end{verbatim}
% \vspace{-1.0em}
  \item \textbf{Figs.~8 (frequency scaling).}
  % \vspace{-0.5em}
  \begin{verbatim}
cd freq && python3 extract_freq_data.py
python3 plot_freq.py
  \end{verbatim}
%\vspace{-1.5em}

  \item \textbf{Figs.~9 (cache sensitivity).}
  %\vspace{-0.5em}
  \begin{verbatim}
cd cache 
python3 extract_l1_cache_metrics.py
python3 extract_l2_cache_metrics.py
python3 plot_l1_mpki.py
python3 plot_l2_mpki.py
python3 plot_l1_norm_exec_time.py
python3 plot_l2_norm_exec_time.py
  \end{verbatim}
  
%\vspace{-1.5em}
  \item \textbf{Fig.~10 (pipeline width).}
    % \vspace{-0.5em}
  \begin{verbatim}
cd width
python3 extract_width_exec_time.py
python3 plot_width_percent_diff.py
  \end{verbatim}
% \vspace{-1.5em}
  \item \textbf{Fig.~11 (LQ/SQ).}
% \vspace{-0.5em}
  \begin{verbatim}
cd load-store-queues
python3 extract_lq_sq_exec_time.py
python3 plot_lq_sq_percent_diff.py
  \end{verbatim}
\vspace{-1.5em}
  \item \textbf{Fig.~12 (branch predictor).}
\vspace{-0.5em}
  \begin{verbatim}
cd branch-pred
python3 extract_branchpred_exec_time.py
python3 plot_branchpred_relative.py
  \end{verbatim}
\vspace{-1.5em}
\end{enumerate}

All generated PDFs appear in their respective subdirectories under \texttt{plot-scripts/}.

%%%%%%%%%%%%%%%%%%%%%%%%%%%%%%%%%%%%%%%%%%%%%%%%%%%%%%%%%%%%%%%%%%%%%
\vspace{-.5em}
% \subsection{Evaluation and expected results -- Part 1}
% Running the experiments should produce VTune reports matching the paper's findings, including:
% \begin{itemize}
% \setlength\itemindent{-2em}
% \item \textbf{Hotspots:} The reports identify the most time-consuming 
%   functions in FEBio (e.g., linear solvers, matrix assembly routines) 
%   and show how execution time is distributed across threads.  

%   \item \textbf{Memory Access:} The reports highlight bandwidth usage, 
%   cache efficiency, and memory stall cycles, which generally indicate 
%   whether the workload is memory-bound or compute-bound.  

%   \item \textbf{Microarchitecture Exploration:} These results provide 
%   a breakdown of front-end, back-end, and core pipeline utilization, 
%    and a summary of the simulation's performance.  
% \end{itemize}

\subsection{Evaluation and expected results -- Part 1}
Experiments should yield VTune reports consistent with the paper, showing:
\begin{itemize}
\setlength\itemindent{-2em}
  \item \textbf{Hotspots:} most time-consuming functions (e.g., solvers, matrix assembly) and thread-level time distribution.  
  \item \textbf{Memory Access:} bandwidth use, cache efficiency, and stall cycles indicating memory- vs.\ compute-bound behavior.  
  \item \textbf{Microarchitecture:} front-/back-end and pipeline utilization with overall performance summary.  
\end{itemize}

%%%%%%%%%%%%%%%%%%%%%%
% \vspace{-1em} 
\subsection{Evaluation and expected results -- Part 2}
% Running the plotting pipeline over the packaged results reproduces
% Figs.~7--12 from the paper:
% \begin{itemize}
% \setlength\itemindent{-2em}
%   \item \textbf{Pipeline stage breakdowns:} stacked and normalized views.
%   \item \textbf{Frequency scaling:} IPC vs.\ frequency and simulation time vs.\ frequency.
%   \item \textbf{Cache sensitivity:} L1/L2 MPKI trends and normalized execution time.
%   \item \textbf{Pipeline width sensitivity:} percent-difference bars relative to the baseline.
%   \item \textbf{LQ/SQ sensitivity:} percent-difference bars relative to the baseline.
%   \item \textbf{Branch predictor sensitivity:} relative execution time vs.\ baseline.
% \end{itemize}
% \vspace{-.25em}
The plotting pipeline over packaged results reproduces Figs.~7--12:
\begin{itemize}
\setlength\itemindent{-2em}
  \item \textbf{Pipeline stage breakdowns:} stacked and normalized.  
  \item \textbf{Frequency scaling:} IPC and runtime vs.\ frequency.  
  \item \textbf{Cache sensitivity:} L1/L2 MPKI and normalized runtime.  
  \item \textbf{Pipeline width sensitivity:} percent-difference vs.\ baseline.  
  \item \textbf{LQ/SQ sensitivity:} percent-difference vs.\ baseline.  
  \item \textbf{Branch predictor:} relative runtime vs.\ baseline.  
\end{itemize}
Minor numerical deviations may appear if re-running gem5 due to simulator version drift or host system variability.

%%%%%%%%%%%%%%%%%%%%%%%%%%%%%%%%%%%%%%%%%%%%%%%%%%%%%%%%%%%%%%%%%%%%%
%\subsection{Experiment customization}
% Advanced users can edit or extend sweeps by modifying:
% \begin{itemize}
%   \item \texttt{run-scripts/run\_other\_configs.sh} and \texttt{run-scripts/run\_micro\_configs.sh} (which call the parameterized launchers).
%   \item Per-feature launchers in \texttt{run-scripts/} (e.g., \texttt{febio\_run\_freq\_exps.sh}, \texttt{febio\_run\_l1\_exps.sh}, \texttt{febio\_run\_l2\_exps.sh}, \texttt{febio\_run\_branch\_pred\_exps.sh}, \texttt{febio\_run\_micro\_exps.sh}).
% \end{itemize}

% \subsection{Experiment customization — Part 1
% } 
% \vspace{-.5em}Users can add their own FEBio models by placing \texttt{.feb} input files 
% into the \texttt{bin/} directory and re-running the workflow scripts. 
% Analysis type can be changed by modifying \texttt{scripts/run\_all.sh}.  

%%%%%%%%%%%%%%%%%%%%%%%%%%%%%%%%%%%%%%%%%%%%%%%%%%%%%%%%%%%%%%%%%%%%%
\vspace{-1em}
% \subsection{Notes}
% \vspace{-.5em}
% % FEBio binaries ship only with x86-64 support. Arm-based Macs can run using Rosetta 2,  but profiling is only supported on Intel hardware. A future Arm-native FEBio port exists  (experimental, using Panua Pardiso) but is not included here. 

% % If re-running simulations, allocate long wall-clock time and sufficient storage. For most reviewers we recommend mode (B): use the packaged results and run only the plotting steps. For mode (A), Ubuntu LTS with $\geq$8 CPU cores and 64\,GB RAM is sufficient. \vspace{-.75em}

% FEBio binaries ship only with x86-64 support. Arm-based Macs can run via Rosetta~2, though profiling is supported only on Intel hardware. An experimental Arm-native port (using Panua Pardiso) exists but is not included here. If re-running simulations, allocate long wall-clock time and sufficient storage. For most reviewers we recommend mode~(B): use the packaged results and run only the plotting steps. For mode~(A), Ubuntu LTS with $\geq$8 CPU cores and 64\,GB RAM is sufficient. %\vspace{-.75em}

\vspace{1em}

%%%%%%%%%%%%%%%%%%%%%%%%%%%%%%%%%%%%%%%%%%%%%%%%%%%%%%%%%%%%%%%%%%%%%
\subsection{Methodology -- Part 1 %(pointer) — Part A
}
%\vspace{-.5em}

We profile FEBio’s \emph{Stage 2} (solver) using VTune Microarchitecture Exploration, capturing instruction-level efficiency, memory/cache behavior, and pipeline stalls; models are drawn from the FEBio test suite, with an \emph{eye} case study for stress-testing. 
% \subsection{Methodology}
% Submission, reviewing and badging methodology:
% \begin{itemize}
%   \item \url{https://www.acm.org/publications/policies/artifact-review-and-badging-current}
%   \item \url{https://cTuning.org/ae}
% \end{itemize}

%\end{document}

\end{document}